# STRATEGIC PLAN 2021-2030
## FOR ASTRONOMY IN THE NETHERLANDS

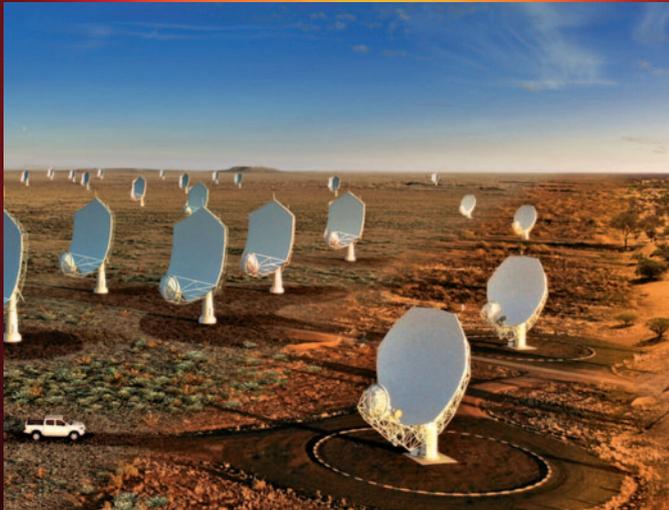
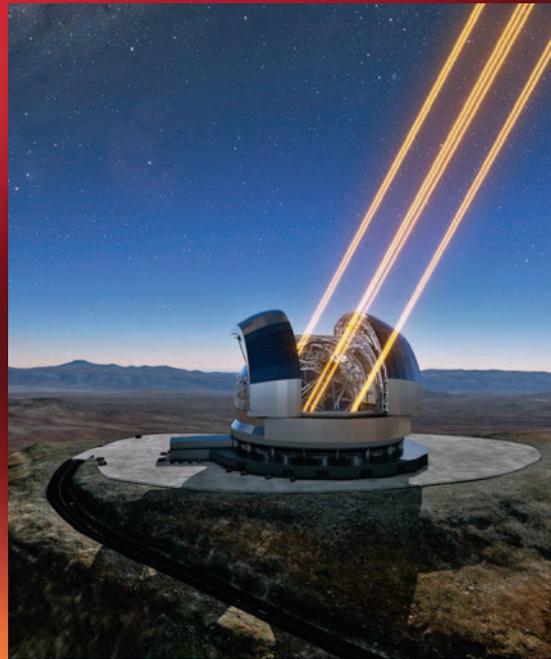
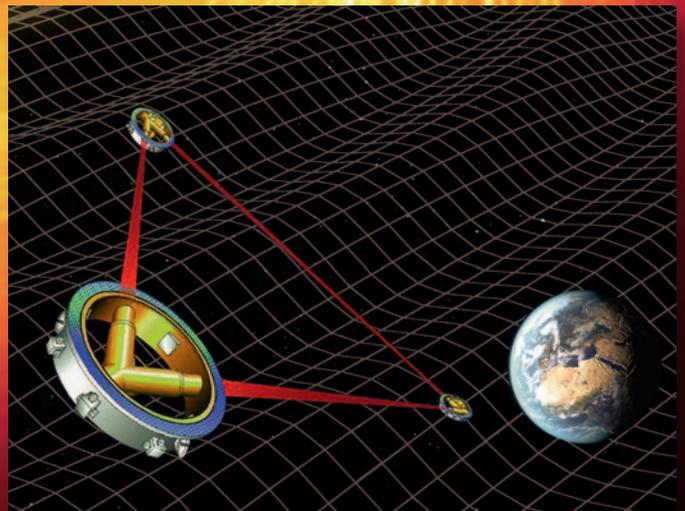

RAAD VOOR DE ASTRONOMIE
ON BEHALF OF NOVA, SRON, ASTRON

R.A.M.J. WIJERS
K.H. KUIJKEN
M.W. WISE (eds.)

SEPTEMBER 2022

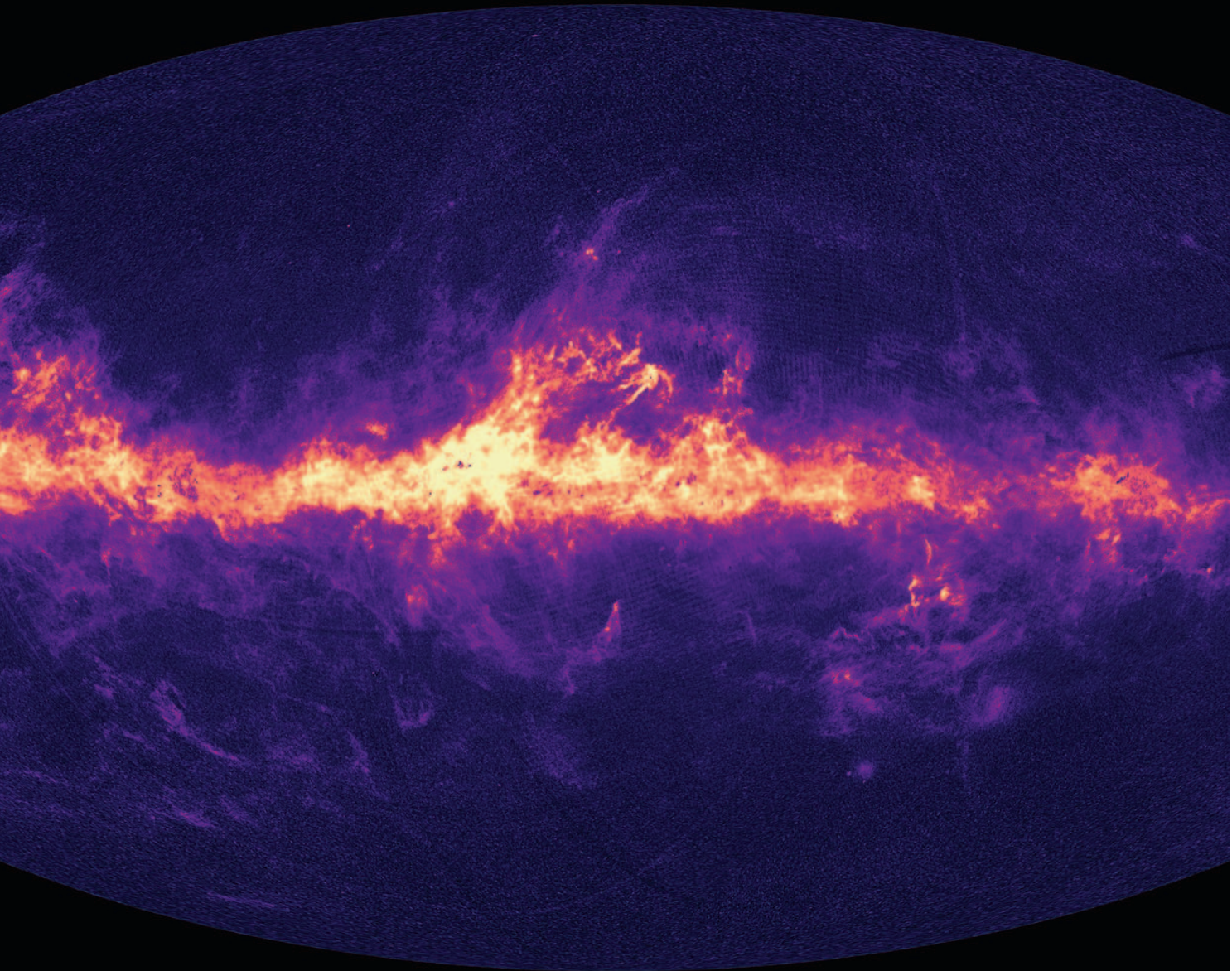

▲ The distribution of dust across the sky as imaged by ESA's Gaia mission.

◀ Cover image: Our three top facility priorities, LISA, ELT, and SKA shown as jointly observing a complicated object, the black hole at the centre of the galaxy M87.

# TABLE OF CONTENTS

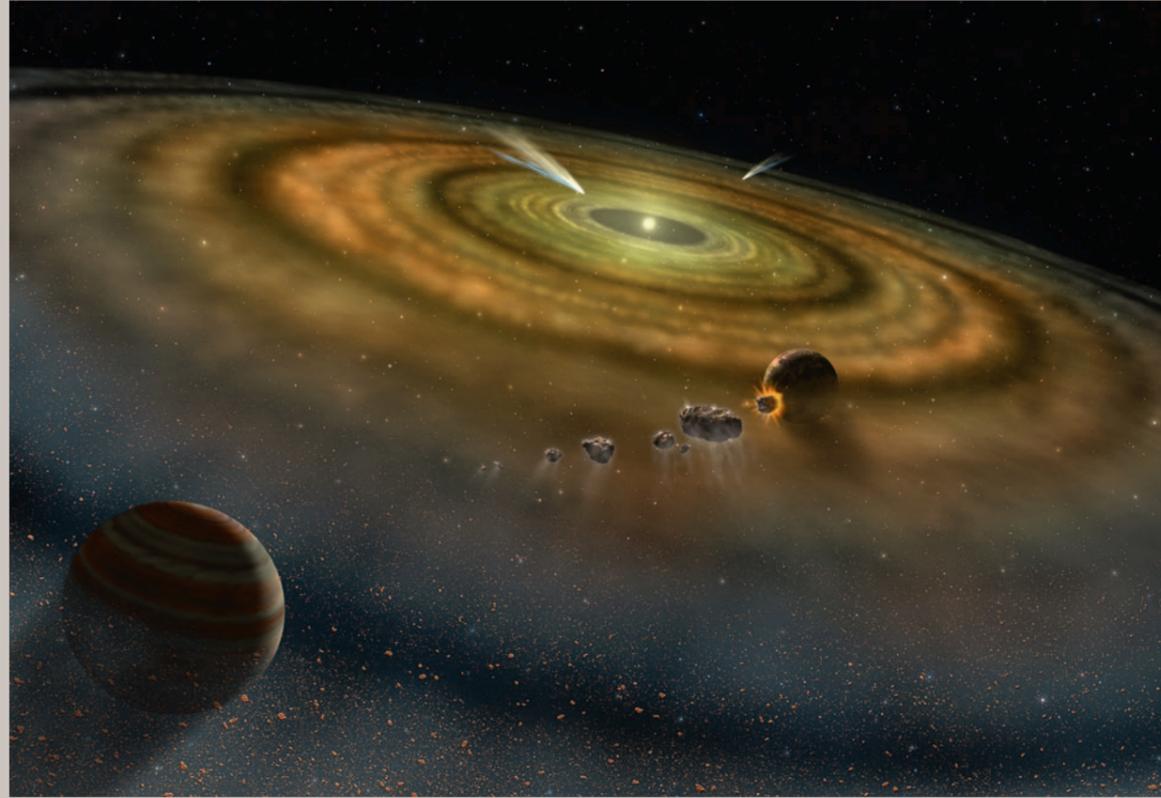

# EXECUTIVE SUMMARY

Astronomy is the quest for understanding the very fabric of our Universe and everything in it, and our history and place among its stars and galaxies. Ancient as humanity itself, it has become deeply rooted in our culture as it is passed through the generations. The first astronomers noticed the regular motions of stars and planets in Earth's sky, prompting attempts at understanding the forces that drive them. Progress has been enormous since then, as we discovered how vast and rich our Universe is beyond our Solar System, and the large changes it underwent in the 13.7 billion years since the Big Bang.

In the coming decade, our quest will focus on inspiring topics: finding the earliest stars and galaxies and understanding how they formed; measuring the finest ripples in spacetime to understand its most extreme events; and exploring other planetary systems, possibly finding clues to what makes life itself. We will also encounter surprises, which will inspire new questions and new ways of addressing them for the equally inspired astronomers of the next generation. This quest is a long game that proceeds in many small steps and the occasional longer stride, but always needs a long-term perspective. Good ideas are the fruit of long periods of training, collaboration and thought, and top observatories need decades to develop, build, and exploit. The environment that attracts and fosters the required talent must have stable base funding and offer opportunities to lead world class efforts. As a small country, we need to set a few strategic priorities from the many available options, based on what we excel in, and execute these optimally in a nationally concerted effort. Prioritization must consider our scientific interests, strengths, and existing technical expertise and that of our industrial partners. The complexity and cost of top facilities require joint efforts on the international scale, and thus international coordination and alignment of priorities. This decadal plan presents the choices we made for the period until 2030, and due to the long timescales even looks well beyond that.

Our first priority is to invest in a sustainable community of excellence. This requires that the funding of our NOVA top research school becomes structural. It also requires investing in expertise in compute- and data-intensive astronomy, in equity, inclusion and diversity and in energy sustainability. Our other priority is to invest in the development and building of three select global facilities that best support the focal topics of our scientific quest: (I) instrumentation for the best and largest optical/infrared telescope in the world, ESO's Extremely Large Telescope, exploring distant planets and the history of galaxies from the earliest times until the present; (II) participation in ESA's gravitational-wave observatory LISA, exploring the violent mergers of black holes;

(III) the Square Kilometre Array and future technology for global radio astronomy, exploring a wide range of topics from pulsars and fast radio bursts to cosmic dawn.

With such stability and specific investments, we can continue to provide global leadership in astronomy and attract young talents to join that quest. We will also keep up our vigorous outreach programme to help captivate many others to science in general and inspire everyone. Beyond sustaining our own research, the talented people we train in complex problem solving and data science will help address many societal problems of similar scale and complexity; our industrial partners will be stimulated to push technology to new limits.

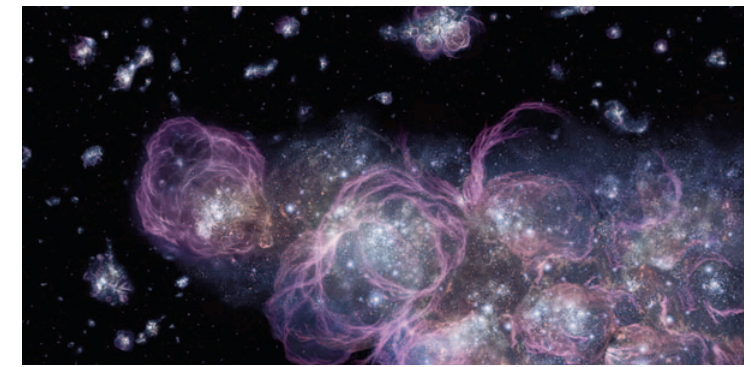

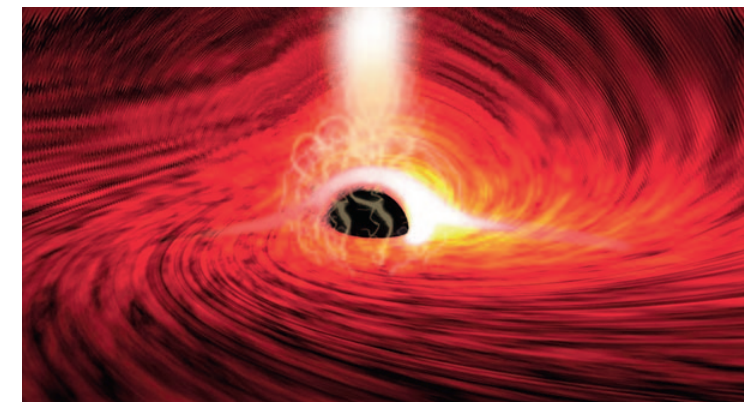

▲ *In the coming decade, we will explore how the first stars formed in the earliest galaxies (top above), how planets like Earth form in young solar systems (top left page), and how black holes energise the most extreme processes in the Universe (above).*

▼ *All-sky map of stars in the Milky Way, as measured by the Gaia satellite.*

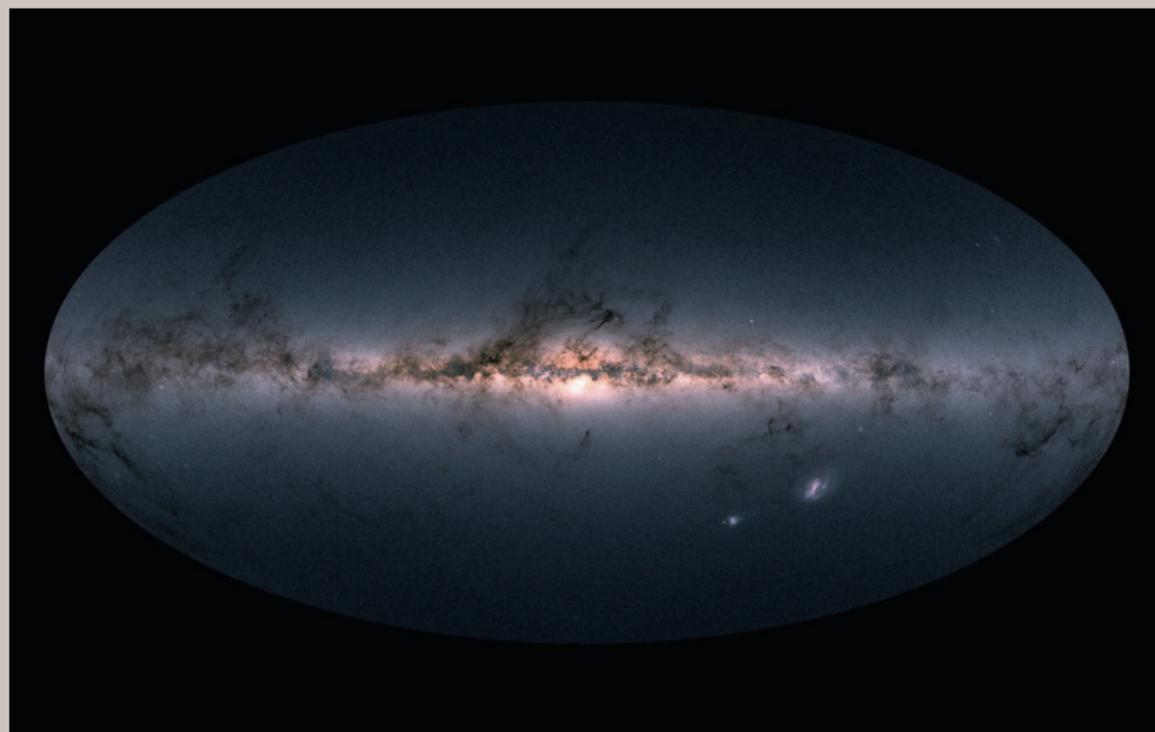

# INTRODUCTION

This report is the decadal strategic planning document issued on behalf of astronomy in The Netherlands by the Astronomy Council (Raad voor de Astronomie, RvdA). It follows our tradition of national coordination of science and facilities established in the middle of the last century, and represented via formal strategic reports since 1986. Its purpose is to organize the relevant Dutch scientific and technological expertise into a number of clear, long-lasting lines of research and contributions to facilities that optimize our ability to do the best science and have international impact.

The decadal reports present concrete plans for science and facilities for the coming decade, in this case 2021-2030. A midterm update will be used to sharpen the plans for the second half of the decade. These plans outline the resources required to realize the ambitions we formulate. These reports also preview an increasingly long period after their nominal horizon, in the current case up to 2050, due to the very long planning cycles of our flagship facilities and the Inter-Governmental Organisations (IGO's) that run and support them.

The core of Dutch astronomy research consists of a partnership between the NWO institutes ASTRON and SRON and the university astronomy departments allied in the NOVA research school. Together they have enabled the Netherlands to continue to do breakthrough research on well selected research themes, even as the field of astronomy became a Big Science organised increasingly around large international facilities. NOVA won top research school status from 1999 and used these targeted funds to invigorate the research environ- ment at the universities and to enable participation in the ESO VLT and the ALMA instrumentation programmes. This participation gave access to guaranteed observing time on these frontline observatories and was the stepping stone for NOVA to become involved in first genera- tion instruments for the ELT. ASTRON focuses on radio astronomy and is a world leader in the development of new instruments for radio telescopes, such as the International LOFAR Telescope with its core in Drenthe, and the world-wide SKA. SRON is the Dutch expertise institute for space research. Its astrophysics programme has world leadership in X-ray and far-infrared instrumental development for ESA missions such as ISO-SWS, Herschel/HIFI, XMM-Newton, and Athena. Important strategic decisions that affect Dutch astronomy (such as this decadal plan) are coordinated in the RvdA, a body that brings NOVA, ASTRON and SRON together.

The report is organized as follows: in chapter 1, we review the state of astronomy and formulate our scientific ambitions for the coming decade. In chapter 2, we review the state of our community and the context in which it operates and examine what is needed to move it forward robustly and sustainably. In chapter 3, we map out the select priorities that follow from our analysis and the funding profile required for them. Finally, in chapter 4, we look back to the results of the previous decade, to cross-check the realism of our planning process, and preview the decades after 2030, to examine what foundations we must lay already now to continue our success beyond the current planning period.

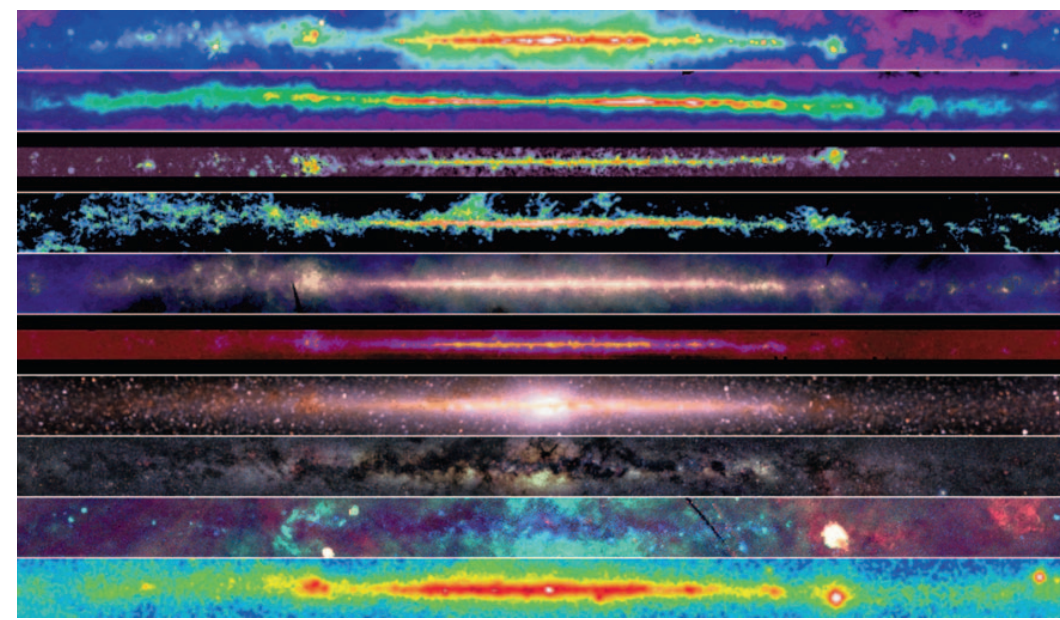

▲ *Images of the Milky Way band across the sky in 9 different wavebands, from radio waves (top) to gamma rays (bottom).*

▼ *The central 1000 light years of the Milky Way imaged in radio waves by the MeerKAT telescope (frequency 1.4 GHz). The bright source at the centre is Sgr A\* and the Arcs, trails and supernova remnants around it show many different signs of energetic processes in this region such as the central black hole and remnants of exploded stars. The different colours indicate different shapes of the radio spectra of the sources.*

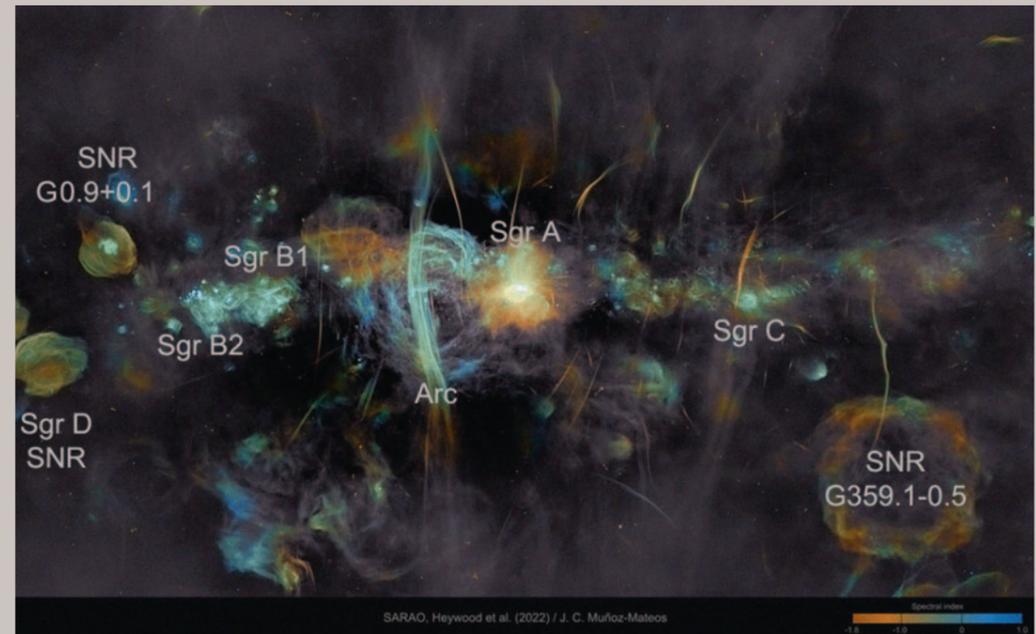

# 1  SCIENCE VISION

When looking back over the past century of astronomy we cannot escape being astonished at the progress made. A century ago, it was only just becoming clear that our Milky Way is one of countless other galaxies, but little was known about our Galaxy itself, or our location within it. Pluto had not yet been discovered, and the origin of the Sun and the Solar System as well as the stars was a mystery. The expansion of the Universe was yet to be recognised.

## 1.1  THE BIG QUESTIONS FOR THE 21ST CENTURY

The past century has seen an amazing number of discoveries: the nuclear reactions that power stars, the cold gas clouds out of which stars form, the life cycles of stars, the outer Solar System and the Kuiper Belt, the cosmic background radiation, large-scale structure of galaxies, white dwarfs and neutron stars, stellar-mass and supermassive black holes, quasars, dark matter and dark energy, exoplanets, interstellar molecules, gravitational waves, cosmic rays, etc. etc. Astonishingly, the pace of discovery is far from saturated! Important open questions on the nature and origin of planetary systems, galaxies, and the cosmos are within reach of major breakthroughs.

Technology and state of the art observational and computational facilities are major drivers of our astronomical research, together with theoretical insights, simulations, and imagination. Astronomers have been quick to apply new detector and telescope techniques as they appeared but have also successfully pushed key technologies themselves (such as interferometer arrays for radio astronomy, or satellites for infrared, ultraviolet, X-ray, and gamma-ray observations). These

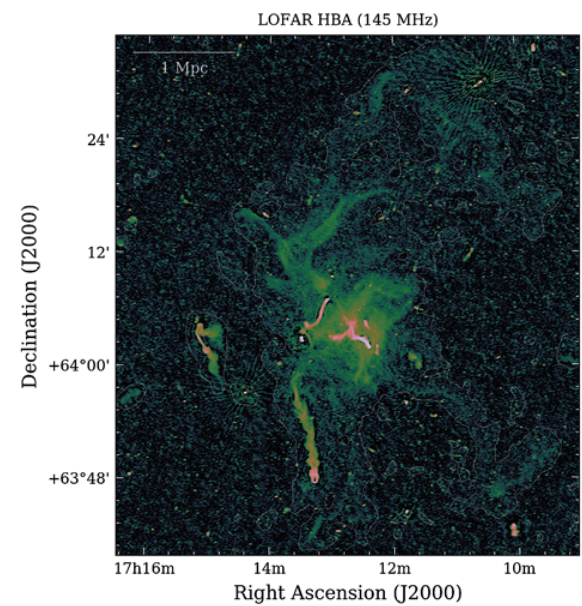

▲ *Galaxy clusters like A2255 are the most massive structures in the Universe. This LOFAR observation reveals the gas between galaxies and clusters of galaxies, which shines due to the magnetic fields and high-energy particles in it.*

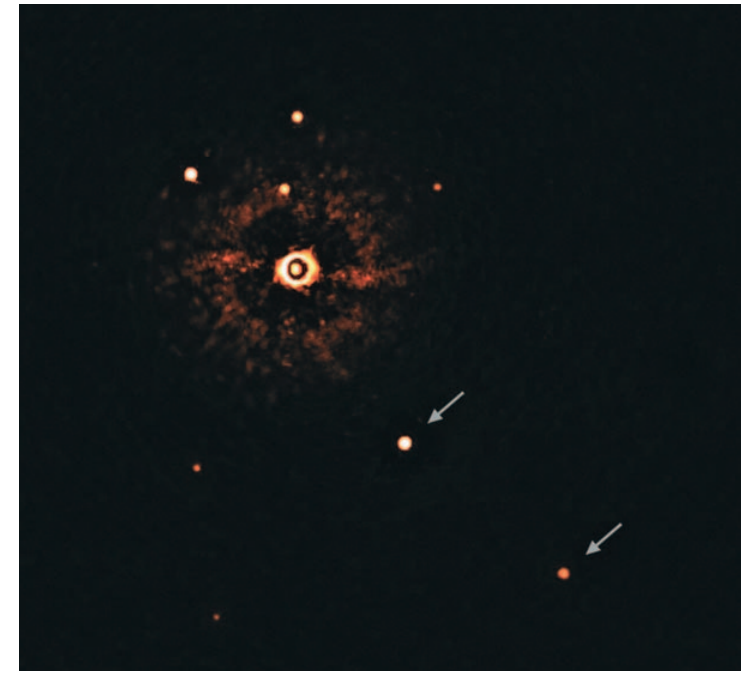

▲ *First ever image of a young, Sun-like star accompanied by two giant exoplanets (indicated by arrows), taken by the SPHERE instrument on ESO's Very Large Telescope. The system is located about 300 light-years away and known as TYC 8998-760-1.*

hand-in-hand developments continually put new research questions within reach.

These are exciting times. In this decade, ESO's Extremely Large Telescope (ELT) with a 39-m primary mirror will come online, with other similarly large telescopes possibly following later. The ELT is designed to deliver maximally sharp images with which it will be possible to separate the light of Earth-like planets orbiting neighbouring stars from the glare of their sun and study their atmospheres. The recently launched James Webb Space Telescope (JWST), the long-awaited successor to the Hubble Space Telescope, will bring the formation of the first galaxies, 13 billion years ago, into view. The first phase of the Square Kilometre Array (SKA) will transform radio astronomy, galvanising the study of cosmic dawn and the epoch of reionisation, and of pulsars. The new field of gravitational wave astronomy, where we will build LISA and Einstein Telescope (ET) to study the most energetic and rarest events in the Universe, the mergers of black holes, is part of the exciting growth of multi-messenger astronomy. The Euclid mission will be launched to map the growth of large-scale structure over the past 8 billion years, in an effort to elucidate the nature of dark matter and dark energy.

Combined with underpinning theoretical studies, advanced computer simulations, and great advances in data science, these new facilities will enable us to address the most fundamental questions about the Universe as well as our place in it: how did the Universe come to be the way it is, what is the physical nature of its ingredients, how do galaxies, stars and planets form, evolve, and die, and ultimately, what are the implications for the origin and prevalence of life?

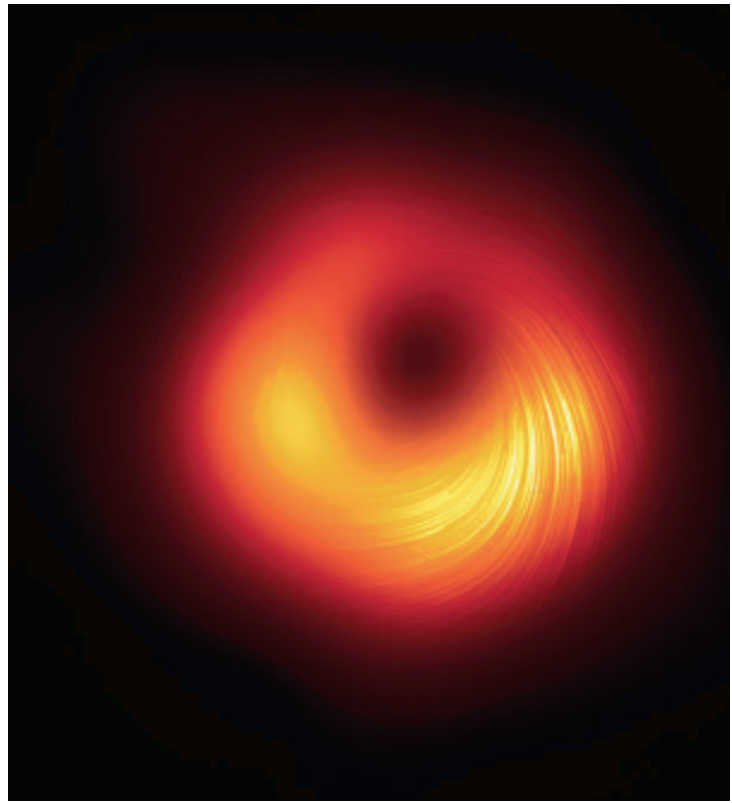

▲ *The supermassive black hole at the centre of the galaxy M87 imaged by the Event Horizon Telescope. The dark centre is the shadow of the black hole, The brighter regions are gas swirling around it, lensed by the black hole's gravity. The lines indicate polarisation of the light, due to the way very energetic particles radiate.*

Our Dutch astronomy community is prominent in this global effort, with leading roles in a number of key research areas and well-developed plans and ambitions for using important new facilities that will see first light in the coming decade. We briefly describe these below, grouped in the three main themes of Dutch astronomy research. These science questions also provide the context for the strategic investment plans in Section 3, whose scientific harvest will be the foundation for astronomy post-2030.

## 1.2 SCIENCE QUESTIONS FOR THE NEXT DECADE

### 1.2.1 EVOLUTION OF OUR UNIVERSE AND ORIGIN OF THE MILKY WAY

Galaxies connect the largest and smallest structures of the Universe. They dominate our view of the faint sky, they are the environment where stars and planets are born and die, and they delineate the large-scale structure of the cosmos. Faraway galaxies also provide a view of the younger Universe, allowing astronomers to study its history directly. In the coming decade we will be able to look further back in time, allowing us to see the first generation of stars and galaxies. Witnessing this "dawn of the Universe" is essential for our understanding of this important phase in cosmic history. More clues about this dawn come from scouring the local Universe for very old metal-poor ('pristine') stars that provide a fossil record of the earliest stages of galaxy formation and allow us to do 'Galactic archaeology'. Combining this with advances in numerical simulations we will obtain a more complete theory of galaxy and structure formation, where gas dynamics play an important role. Understanding how the physics of gaseous matter shapes the observable Universe is also essential to interpreting the information encoded in the positions and shapes of galaxies. A new generation of wide field imaging and spectroscopic surveys will map these with unprecedented accuracy, with the aim of shedding light on the nature of dark matter and dark energy.

We can divide our research for the coming decade into three intimately connected main themes:

**What happened during the dawn of the Universe?** The Universe started off hot and nearly homogeneous, growing colder and clumpier as it expanded. As the clumps became denser, the first galaxies and stars formed in them, ultimately leading to the complex Universe that we see around us today. The details of this important transformation are largely unknown because the physical conditions were very different to those in the current Universe. To understand what happened in the first billion years after the Big Bang, we need to know the properties of those first stars and galaxies. Supermassive black holes are believed to play an important role in regulating star formation in galaxies, but how they formed and influenced galaxy formation and evolution remains a big outstanding question. Related to all this is reionization, the era in which the energy produced by the first stars and galaxies made the hydrogen in our Universe go back from neutral to ionised. What is the basic timeline and which sources drove the process? New revolutionary facilities starting to operate within this decade, such as the JWST, ELT, Euclid and SKA, will allow us to study the early phases of galaxy evolution directly, while VLT, ALMA and LOFAR will continue to play important roles. These studies of distant galaxies are complemented by local studies of pristine stars.

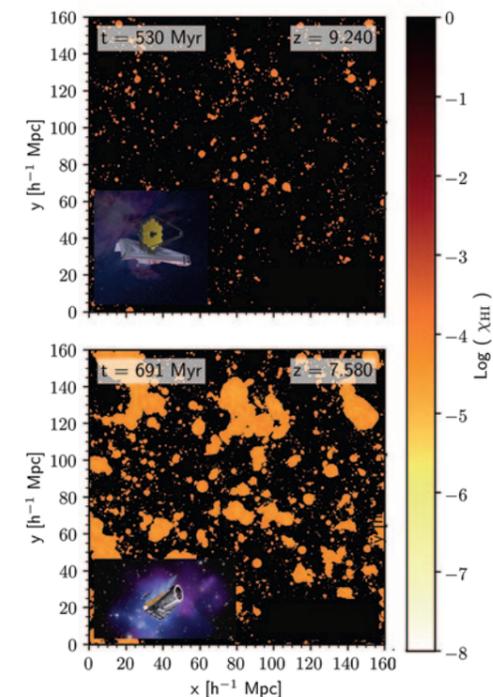

▲ *The ASTREUS simulation shows how reionization in the Universe proceeds from redshift 9.2 (top) where the first light has only created small ionised regions (shown in orange) to 7.6, where much more is already ionized. JWST will discover the first galaxies shining at the highest redshift, whereas Euclid will map the brightest ones towards the end of reionization (adapted from Hutter et al. 2020)*

New spectroscopic surveys carried out by 4MOST and ING/WEAVE will allow us to study these stars identified in the Milky Way using the Gaia mission, whereas the ELT will extend such studies to other galaxies as far as the Virgo cluster.

**What is the nature of the main constituents of the Universe?** 95% of the content of the Universe is dark matter and dark energy, which challenge our understanding of both cosmology and fundamental physics. Dark matter outweighs normal matter six-to-one and drives the formation of large-scale structure in the Universe. We are yet to find explanations for its physical nature. An even bigger puzzle is understanding what is the dark energy that causes the accelerated expansion of the Universe. We will make progress in the coming decade using huge new data sets from multi-object spectrographs and wide-field imagers; they will show how dark matter, stars and gas are distributed in the Milky Way and other galaxies, and how these constituents influence each other. Finally, we will map the evolution of structure formation with dedicated wide-area surveys, allowing us to test new theories of dark energy and gravity. The WEAVE and 4MOST surveys, in combination with Gaia and Euclid, will allow us to make detailed maps of substructures in our Galaxy and distant galaxies via strong lensing, which are potentially associated with dark matter clumps in the Milky Way halo. In parallel, the Euclid galaxy survey over 40% of the sky will test the current cosmological model with unprecedented precision.

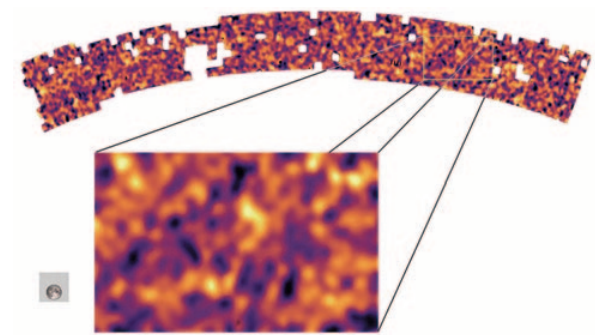

▲ *A map of the dark matter distribution, produced by the KiloDegree Survey and derived from weak gravitational lensing. The degree of clustering (clumpiness) of the matter is a sensitive probe of the competition between gravity on the largest scales, and the expansion of the Universe. The full moon is shown to the same scale as the inset.*

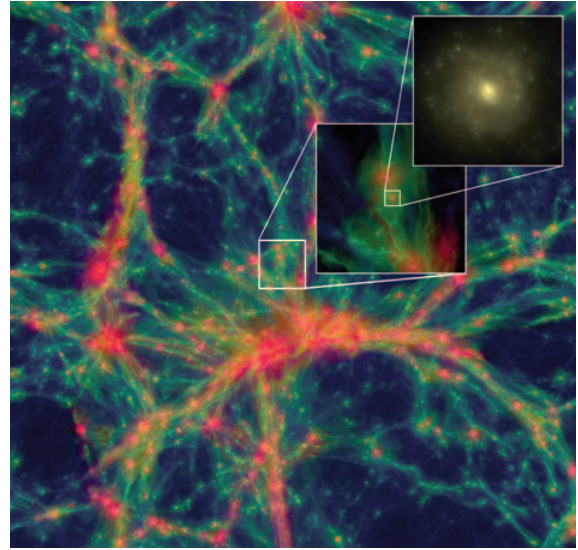

▲ *The EAGLE project is a campaign of large-scale hydrodynamical simulations of the Lambda-Cold Dark Matter Universe. In this three-scale zoom, the largest image shows the cosmic web of dark matter and gas clumping together, and the smallest a newly formed galaxy as it might appear to us in a powerful telescope.*

**How did galaxies assemble and evolve?** To make sense of the wide range of multi-wavelength data we rely on a cosmological framework and on understanding how galaxies form and evolve. The former is largely in place, but despite progress over the last decade the latter still has major open questions. We are far from a complete theory of galaxy formation with strong predictive power. We do not understand the processes that halt, start, or sustain star formation well. Studying the complex interplay of galaxies and their multi-phase environment will be fundamental to investigating these processes. Particularly, the role of dust in galaxy evolution is poorly addressed in galaxy models, even though its role was very important in the first half of cosmic history. Finally, a key ingredient to galaxy models is the presence of active nuclei, but how in reality they regulate star formation is very unclear. Over the next decade, we expect major progress from advances in hydrodynamical simulations and new high-quality observations over a wide range in wavelength, from X-ray (XRISM, followed in the 2030s by Athena), optical and near-infrared (VLT, ELT, Euclid) and infrared/submm (JWST and ALMA) through to radio (LOFAR/SKA).

### 1.2.2 ORIGIN OF STARS AND PLANETS AND OUR PLACE IN THE UNIVERSE

Since the turn of the millennium, we have come to know that planetary systems around other stars are very common – even the norm. At the same time, we found the diversity and complexity of exoplanets and exoplanetary systems to be very large. In the Milky Way alone there must be billions of planets with a composition and climate not unlike Earth: an unknown fraction of these may be habitable. New detailed studies, peering into the hearts of stellar nurseries far out into our Galaxy, allow us to witness the births of new stars and their planets and study the complex physical and chemical processes at play. Understanding the link between these formation processes and the exoplanet population is a major goal for the coming decade. At the same time the prospect of studying exoplanets directly through imaging and transit observations promises a wealth of information on their compositions and atmospheres. Specific questions we will address include:

**How do planetary systems form around newly born stars?** What determines the architecture of planetary systems? What determines the size and composition of planets? Studying how planets of different masses and in different stages of formation influence the appearance of protoplanetary

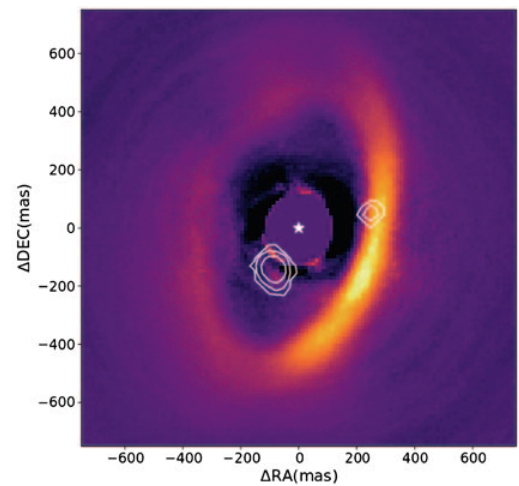

▲ *Composite image of PDS 70, a young star with a circumstellar disc and planets. The colour image shows the disc as seen by the SPHERE instrument on the VLT. The contours of H$\alpha$ show the locations of two planets, betrayed by the fact that matter is falling onto them (from Haffert et al. 2019).*

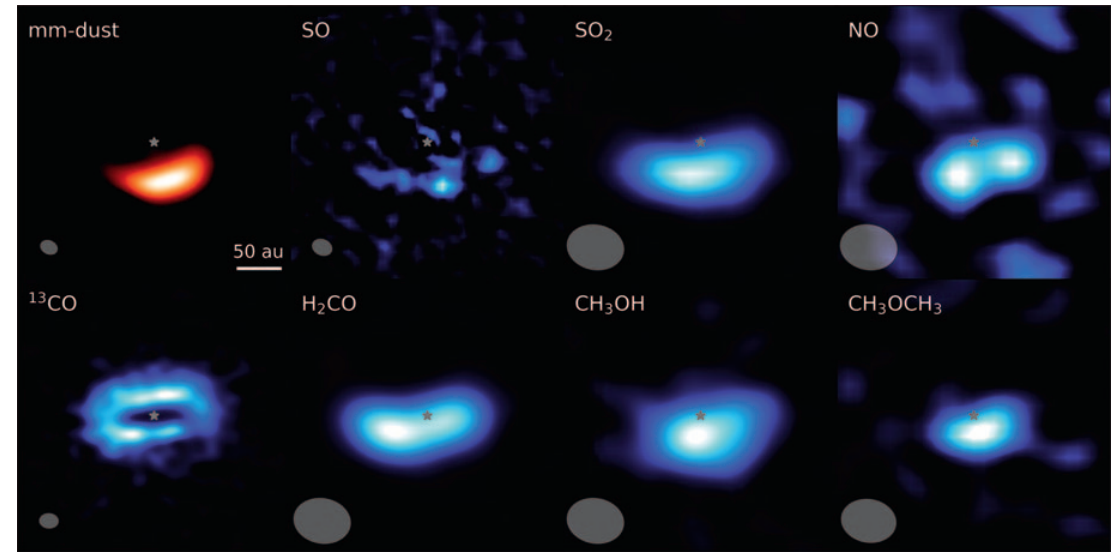

▲ *Gallery of images of a planet forming disc in the source IRS48. The millimetre dust grains (top left) are concentrated in a so-called 'dust trap' in the Southern part, whereas the CO gas emission (bottom left) shows a complete ring. Recent ALMA observations show complex organic molecules, the precursors of the building blocks of life, co-spatial with the dust. This correlation shows that these complex molecules originate from sublimation of the ice layers on the dust grains in the dust trap. These icy grains are most likely inherited from the dark, cold cloud in which the star was formed, as the current disk is too warm to have significant ice chemistry.*

disks - their birth environments – is central to this endeavour.

**What is the diversity of exoplanetary systems?** What are the bulk compositions and atmospheric makeups of different classes of exoplanets? What are their climates like? It is vital that we survey the physical, chemical, and geological processes in exoplanets and their atmospheres, to recognize connections to their formation and evolution. In the case of temperate Earth-like planets, we need to assess potential habitability and search for signatures of biological activity.

**How do complex molecules form?** What is the inventory of prebiotic molecules in space? How are they incorporated in planetary environments? We will first combine state-of-the-art telescopic observations, laboratory experiments, and detailed modelling to focus on which molecules can be found and where. With that information, we can then unravel the formation schemes of molecules in space and understand how they can form so effectively under the harsh conditions that exist there.

**What role does the Galactic context play?** How do stars and planets form in different chemical surroundings? How do massive stars shape their environments? Massive stars produce the bulk of the chemical elements in the Universe, which are released into the interstellar medium and supply the essential ingredients for the formation of Earth-like planets and life. The next generation of telescopes will unlock access to the formation of the most massive stars in other galaxies and galactic environments. Understanding this formation in turn provides important links to the research on galaxies (massive stars are thought to be good proxies for the first stars to form in the Universe) and to high-energy astrophysics (via the study of massive black-hole – black-hole mergers).

Current and near-future facilities such as ALMA/Allegro, JWST, and the ELT, will enable us to make major strides, each with their own unique state-of-the-art capabilities. Space missions such as ARIEL and PLATO will complement their data sets. With crucial complementary efforts in theory and numerical modelling, we will guide and interpret these increasingly complex observations. Laboratory experiments provide basic spectroscopic information and simulate the chemical processes under interstellar conditions; they are an integral part of our research, and one in which we are particularly strong. ALMA is currently already galvanizing the

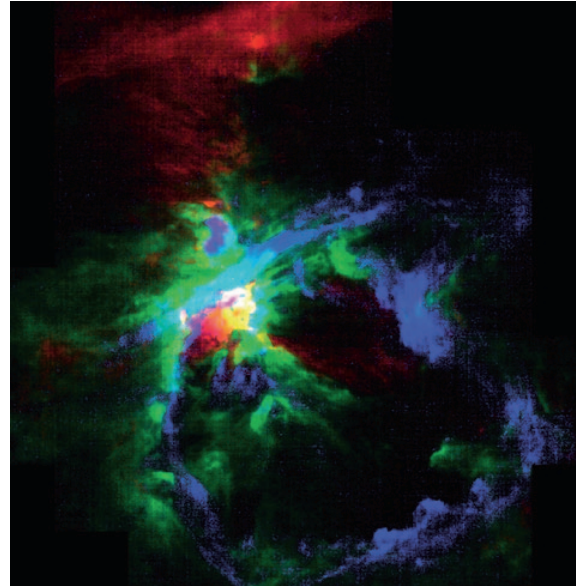

▲ *The complicated distribution of gas in part of the Orion Nebula, a very nearby star forming region. The blue and green colours emphasize a bubble blown by the wind from a massive star and indicate how significantly such young stars shape star forming regions. Infrared data from SOFIA and the Herschel and Spitzer satellites (from Pabst et al. 2019).*

study of the gas and dust properties of protoplanetary disks, enabling us to catch planet formation in the act. JWST will for the first time open the door to detailed atmospheric characterisation of extrasolar planets, down to temperate planets between Earth and Neptune in size. The ELT, including first-light instruments METIS and MICADO, will revolutionize every aspect of star and planet research in the next decade and beyond, including investigating their habitability and the search for life.

### 1.2.3 SPACE AND MATTER UNDER EXTREME CONDITIONS – THE UNIVERSE AS A LABORATORY

Compact objects - white dwarfs, neutron stars and black holes - present astronomers with extremes of temperature, density, velocity, and magnetic field that are unattainable in Earth-bound experiments. Their strong gravitational fields create these extreme conditions and make compact objects a powerful tool for probing the nature of gravity itself. In this domain of astronomy, we seek to understand the formation and evolution of compact objects, the physical conditions which exist within and close to them, and to deploy them as probes of fundamental physics and cosmology. To do so, we will address these critical and interconnected science questions:

**To what precision can general relativity describe gravity?** Gravitational waves and the direct imaging of the shadow of a supermassive black hole provide us with new opportunities to test gravity. In the next decade we will see exceptional progress in this area through the detection of many gravitational wave mergers by LIGO/Virgo (and later the Einstein Telescope), and the expansion of the detectable gravitational wave frequency regime with LISA. Using the Event Horizon Telescope (which includes ALMA and EVN/JIV-ERIC) we have just revealed the shadow of the Milky Way black hole. We can perform further tests of Einstein's General Relativity via time-resolved observations, especially with key new additions such as the African Millimetre Telescope (AMT). Our studies of Galactic compact binaries will provide other, new and unique, tests.

**What is the nature of dense matter?** The densities of nuclear material within neutron stars exceed those in atomic nuclei. The behaviour of such matter is of fundamental importance to our understanding of numerous physical processes. We will use advanced capabilities of X-ray instruments (such as NICER, NuSTAR, XRISM, Athena, eXTP), as well as gravitational-wave observations to probe neutron stars precisely enough to understand the dense matter inside them.

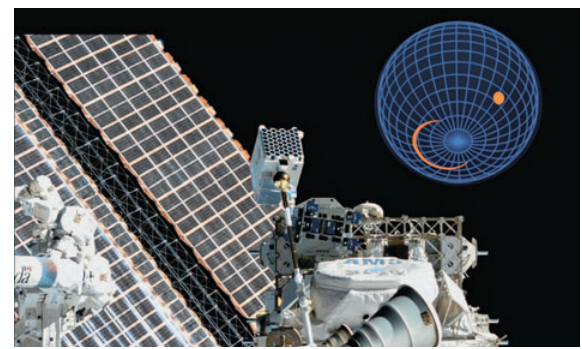

▲ *Careful modelling of X-ray pulse profiles of the pulsar J0030+0451 obtained with the NICER instrument allowed both constraining the properties of the dense matter of which it is made and, quite contrary to expectation, that the magnetic poles from which it emits are neither simple spots not even on opposite sides of the neutron star (After Riley et al. 2019).*

**What is the origin of the elements?** The source of half the elements heavier than iron (the so-called r-process elements, including gold and uranium) remains uncertain. By measuring the event GW170817, the merging together of two neutron stars, we obtained a spectacular demonstration that such an event can create some of these elements. Crucial to this result was our so-called 'multi-messenger' approach: we obtained data both across the entire electromagnetic spectrum ('multi-wavelength') and in gravitational waves. Still, we do not yet know whether they make some, most, or all of these elements. We need more gravitational-wave sources to answer this question and get multi-wavelength data on them, and search large areas of sky for other rare and extreme events (e.g., unusual supernovae). Instruments we need for this are, e.g., the NL-led BlackGEM array, MeerLICHT, NUX, MeerKAT, LOFAR and SKA, and the Swift, Chandra and XMM-Newton observatories. We aim to combine the results with simulations and population synthesis models to finally pin down the origin of every element in the periodic table, connecting with our work on 'Galactic archaeology'.

**How do compact objects form and evolve?** Compact objects mark the end points of stellar evolution, and their numbers and properties tie back to the stars that created them, providing constraints on supernova mechanisms and binary evolution. The different physical properties of white dwarfs, neutron stars and black holes span orders of magnitude in, e.g., mass, temperature, and magnetic field. We get a myriad of observational signatures from their evolution and high-energy processes with observatories such as the VLT and ELT, APERTIF, SKA, Fermi, HESS, ING and INTEGRAL. These potentially allow us to probe the first stars or the build-up of intermediate and supermassive black holes. We will also study time-variable phenomena such as supernovae, gamma-ray bursts, fast radio bursts, tidal disruption events, pulsars and accreting binaries. Combined with indepth stellar evolution and population synthesis calculations these studies will bring us much more complete understanding of how they evolve.

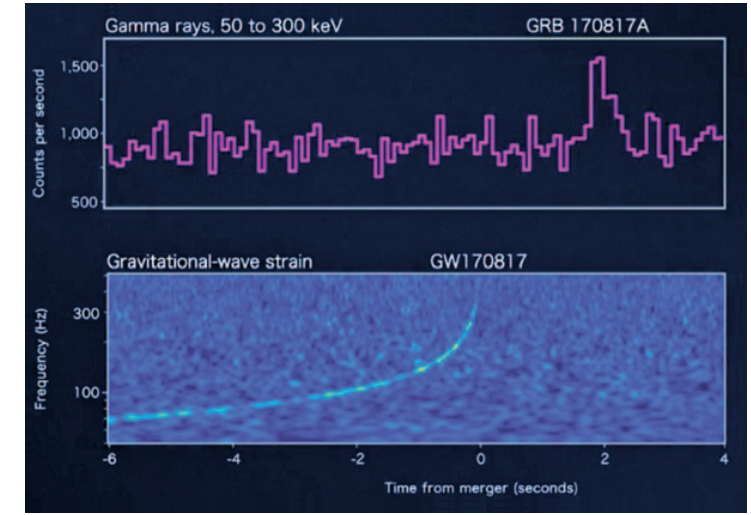

▲ *The true start of the multi-messenger era on 17 August 2017, when gravitational-wave detectors first see the ripples in space-time, or 'chirp', due to a collision of massive objects (bottom), and space telescopes see the gamma-ray signal two seconds later (top), proving that at least one of the colliding objects is a neutron star, ejecting some matter. Later observations in X rays, radio, and optical show the location of the event in a distant galaxy, and prove that these events are where gold and uranium originate.*

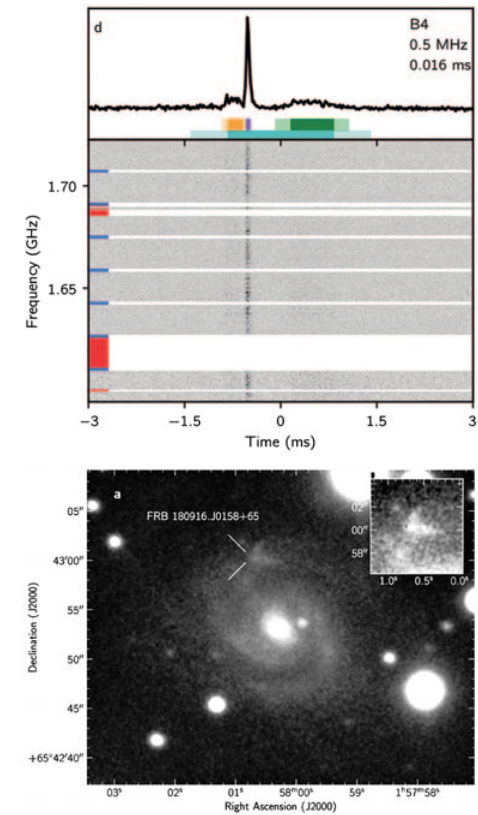

▲ *A Fast Radio Burst shows up as an enigmatic, millisecond-duration, radio transient (top). With great effort, we have managed to pinpoint the location of some, such as the one here on 16 Sep 2018. It turns out to lie within the spiral arm of a distant galaxy, where one typically finds massive stars (bottom). (From Marcote et al. 2020)*

**How do compact objects produce energy and accelerate particles?** Compact objects are powerful astrophysical engines: their strong gravity energises infalling matter. This energised material, sometimes combined with rapid rotation and strong magnetic fields, creates a variety of extreme physical processes, such as discs, jets, outbursts, and outflows. We need highly complex numerical computations to capture the extremes of both physical conditions and scales to understand those processes. These we will combine with new state-of-the-art observations that will capture the most extreme radiation, and non-photonic messengers such as gravitational waves, neutrinos, and high-energy cosmic rays. Facilities that will be key to this are the Pierre Auger Observatory, CTA, KM3NeT, and SKA as well as the above-mentioned X- and gamma-ray instruments.

In this fast-evolving field new opportunities arise at a high rate, thanks to progress in both theory and in multi-wavelength time-domain sky surveys. We can maintain a world-leading position only with access to a large array of facilities, both for discovery and further study of: these extreme objects emit across the full electromagnetic spectrum, as well as in all other messengers (neutrino's, cosmic rays, gravitational waves). Fortunately, our leading position in some facilities gets us access to many facilities led by others (and vice versa). Besides the facilities already mentioned, of future interest for our key questions are the GAMOW satellite and an extended BlackGEM array, and beyond 2030, Athena, the Einstein Telescope and LISA. Critically, we will also require access to supercomputing facilities and an even stronger focus on software creation, data mining and visualisation.

## 1.3 INTERDISCIPLINARY ASTRONOMY

Astronomers in principle study all aspects of celestial phenomena. As the level of detail that we can study becomes finer and finer, astronomy impinges on fields of study such as physics, chemistry, earth sciences, atmospheric science, and even biology. We discuss such areas of overlap in this section. Moreover, our need for increasingly refined and advanced instrumentation and techniques dictates extensive collaboration and overlap with engineering, materials science and signal processing (section 2.5), and computer science (this section).

The level of organisation of our interdisciplinary efforts varies widely. On the one end, they can consist of groups of motivated researchers who have reached across disciplinary boundaries and created long-running associations, sometimes with modest-size facilities attached. Good examples are the Leiden Laboratory for Astrophysics and the Radboud Radio Lab. We also have more recent collaborations that arose from a combination of individual initiative and more programmatic thinking, such as the PEPSci programme for (exo-)planetary science between earth sciences and astronomy and the WARP programme between astronomy and particle physics, both funded by NWO-EW. The Nationale Wetenschapsagenda (NWA; National Science Agenda) funding line, despite not being ideal for fundamental science, did give us an impulse in the areas of black hole research, astroparticle physics and the quest for the origin of life on Earth and beyond; in the latter area we used it to create the Netherlands Origins Centre and local branches like AMCOOL in Amsterdam. Sometimes they consist of more institutional collaboration, such as the inter-institute collaboration on Life in the Universe in Amsterdam, or the astroparticle physics collaborations in all four NOVA universities. Where large facilities are required, such as in astroparticle physics, we have to organise on a scale as large as that of the largest astronomy facilities, via IGOs. Funding for newly emerging interdisciplinary efforts (or even for some older ones) is typically more difficult to assemble than for initiatives in more well-established fields with a long history of organisation. Thus we much need efforts specifically to fund interdisciplinary work, such as those mentioned above.

### 1.3.1 ASTRONOMY AND CHEMISTRY

Astrochemistry has a long and strong tradition in the Netherlands, stimulated over the past decade by our Dutch Astrochemistry Network (DAN), a highly interdisciplinary network combining astronomical and chemical physics expertise in the Netherlands. In it we set the goal of understanding the origin and evolution of molecules in space and their role in the Universe. We received funding for DAN from the Dutch Research Council, NWO, as an integrated and coherent programme combining many elements: astrochemical and astrophysical experiments, quantum chemical calculations, and laboratory spectroscopy of astronomically relevant species. We combined that with an active programme on modelling and observations of astronomical sources. DAN was initiated in 2010 and renewed in 2017. In DAN II we focused on three major astrochemical themes where Dutch astronomy and chemistry have particularly strong expertise and experience as well as access to unique observational or experimental facilities: 1) the Gaseous Universe, 2) the Icy Universe and 3) the Aromatic Universe. The Leiden Laboratory for Astrophysics, a unique facility in the Netherlands, plays a key role in all of these themes. With EU-funded networks such as LASSIE (centred on ices) and EUROPAH (centred on PAHs) we have greatly strengthened the efforts. DAN has been a significant factor in maintaining an active and connected community.

### 1.3.2 ASTRONOMY AND EARTH SCIENCES

Over the past 25 years, astronomy has made great progress in finding and characterizing exoplanets. About 5000 exoplanets have been discovered, and many of them are solid like the Earth. To make progress in understanding the formation, structure and evolution of these rocky exoplanets, knowledge from astronomy needs to be combined with planetary science and the earth sciences.

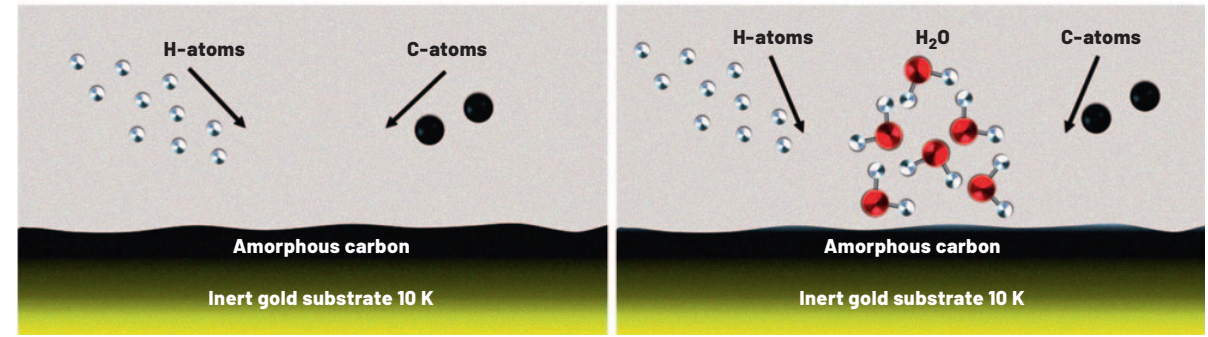

▲ *Visualization of the formation of methane on water poor (left) and water rich (right) dust grains, as probed through laboratory experiments (Quasim et al. 2020).*

Of particular interest are the structure and composition of exoplanet interiors and the composition and dynamics of their atmospheres. Studying rocky exoplanets can teach us about our own Solar System formation; a large variety of exoplanets can be studied at different stages of evolution while we can only directly study our own Solar System at a single evolutionary stage. The interpretation and understanding of exoplanets require the development of new, highly cross-disciplinary and interdisciplinary research approaches. We need to combine astronomical studies with the fields of earth science, biology, and chemistry to provide a coherent framework to understand planetary system formation and evolution – and ultimately answer the question 'Are we alone?'. In 2012, the NWO-funded Planetary and Exoplanetary Science (PEPSci) network was established to foster the collaboration between geoscience and astronomy to establish an integrated, world-leading (exo)planetary research community in the Netherlands, and PEPSci-II was started in 2020.

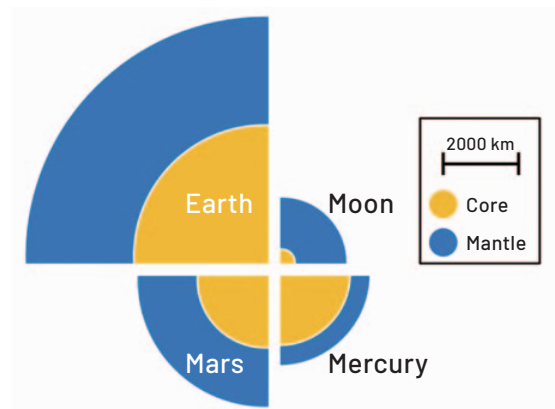

▲ *Studying the internal structure of rocky bodies in the solar system, at the interface of astronomy and earth sciences.*

PEPSci focuses on biomarkers and rocky (exo)-planets. Members of the PEPSci network have also played a leading role in the establishment of the Netherlands Origins Center (origins-center.nl), an NWA-funded Dutch initiative that aims to bring together a very broad range of scientific disciplines to study the origins and evolution of life, planets, and the Universe; and in the Life fellowship programme funded by the EU Horizon 2020 programme and coordinated from SRON (olife-programme.eu).

### 1.3.3 MULTI-MESSENGER ASTROPHYSICS, COSMOLOGY, AND LINKS TO PHYSICS

The interaction between astronomy and physics has always existed but in the past decade has become even stronger. The departments in Amsterdam and Nijmegen have formal collaborations in the GRAPPA (since 2011) and IMAPP (since 2005) institutes. While astronomy at the universities is working together in NOVA, the particle physics community is mainly organised through the National Institute for Subatomic Physics (Nikhef). The astrophysics/physics collaboration is particularly strong around the topics of cosmic rays and gravitational waves and increasingly in dark matter, and in cosmology, black hole research and neutrino astrophysics. We have been fostering and coordinating research in this field for many years through the Committee for Astroparticle physics in the Netherlands (CAN).

**Gravitational Waves and Black Holes:** Our collaboration on gravitational waves in the Netherlands concentrates around a few large international facilities: (I) the Virgo collaboration (of which Radboud Astrophysics and Nikhef/GRAPPA are members); (II) preparations for the LISA space mission selected for launch in 2037; (III) the national involvement in the Einstein Telescope, spearheaded by Nikhef. With the foreseen strong increase of gravitational-wave detections, we expect to develop the field further, using new initiatives and hires in both our physics and astronomy community. In this manner, the physics and astronomy communities each contribute our own strengths to a national synergy in this field.

**Cosmology:** In the Netherlands we have strong physics research groups in theoretical cosmology investigating the very early Universe, as well as astronomy groups studying the late-time Universe through observations and simulations. Increasingly the two communities meet as observational cosmology probes further back in time, and measurements of the cosmological quantities such as dark matter and dark energy get more and more precise. We have started several local initiatives to build a bridge between these communities over the past decade (the Quantum Universe and Fundamentals of the Universe initiatives in Groningen, and the Leiden de Sitter cosmology programme), and our physics colleagues have established a national network for theoretical cosmology. The imminent launch of the Euclid mission, with involvement from astronomers and physicists across the Netherlands, will be a stimulus to strengthen these links even further.

**Cosmic Rays, TeV gamma-rays and Neutrino's:** For more than a decade physicists and astronomers in the Netherlands have been working closely together to investigate the origin of high-energy cosmic rays. Most importantly, we have established the radio detection technique using LOFAR to characterise the radio emission from cosmic rays that is now used worldwide. This has led us into a large effort to install a 3000 km$^2$ radio detection array at the Pierre Auger site in Argentina aimed at the origin of the highest-energy cosmic. We are also involved significantly in TeV gamma-ray observations with HESS and preparing for the future Cherenkov Telescope Array (CTA). Through collaboration with Nikhef, who are involved in KM3NeT, we are also connected with neutrino astronomy, completing the cosmic-ray toolset.

### 1.3.4 COMPUTING, DATA SCIENCE, AND OPEN SCIENCE IN ASTRONOMY

We have been pushing frontier use of computing and numerical algorithms since these became available, both in data processing and modelling, and in theoretical studies of, e.g., fluid dynamics. Over time, the fraction of astronomy that needs to be at the computational frontier to remain competitive or even feasible has grown steadily, to the point that it is now the rule rather than the exception. As a result, we led or co-led significant software projects across all of astronomy: for radio data processing and data mining (WSRT, VLBI, LOFAR), for X-ray spectroscopy (SPEX), for optical/IR surveys (VST/KiDS, Euclid, Gaia, BlackGEM), and for large-scale and/or complex simulations (AMUSE, EAGLE, protoplanetary disks, GR-MHD plus radiative transfer). Increasingly such efforts also count as official contributions to space missions or large surveys, providing Dutch astronomers with another route to access the scientific harvest of such facilities. Also, a greater propor-tion of our work has become data science or data-intensive astronomy, in which the manipulation of large, complex data sets becomes an intrinsic and major part of the problem to be solved. Or it has become computational astrophysics, in which progress in computation and algorithms drives improved theoretical understanding of our Universe.

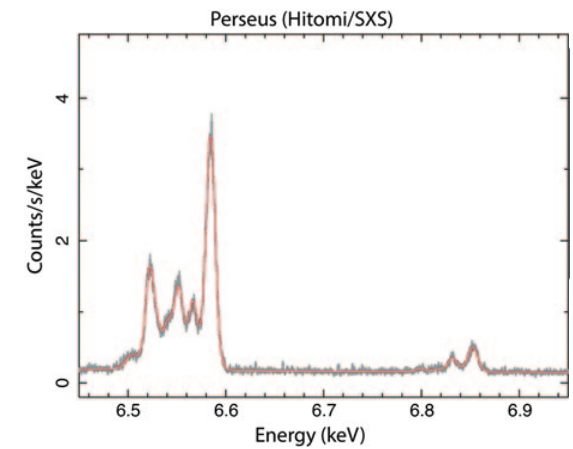

▲ *A detailed part of the X-ray spectrum of the Perseus galaxy cluster observed by the Hitomi satellite, with a detailed model fit (red curve) using SRON's SPEX software package, developed to support plasma physics analysis of data from X-ray missions.*

However, a small number of experts have been leading these efforts, often using temporary project funding. This has made them less well embedded in the structure of our community and has made it hard to provide continuity of expertise and proper long-term curation of the resulting software. We want to ensure that our software, computational astrophysics, and data science efforts contribute to, and even enhance, our leadership position. To that end, we need to make them a strong and structural part of our effort, with the

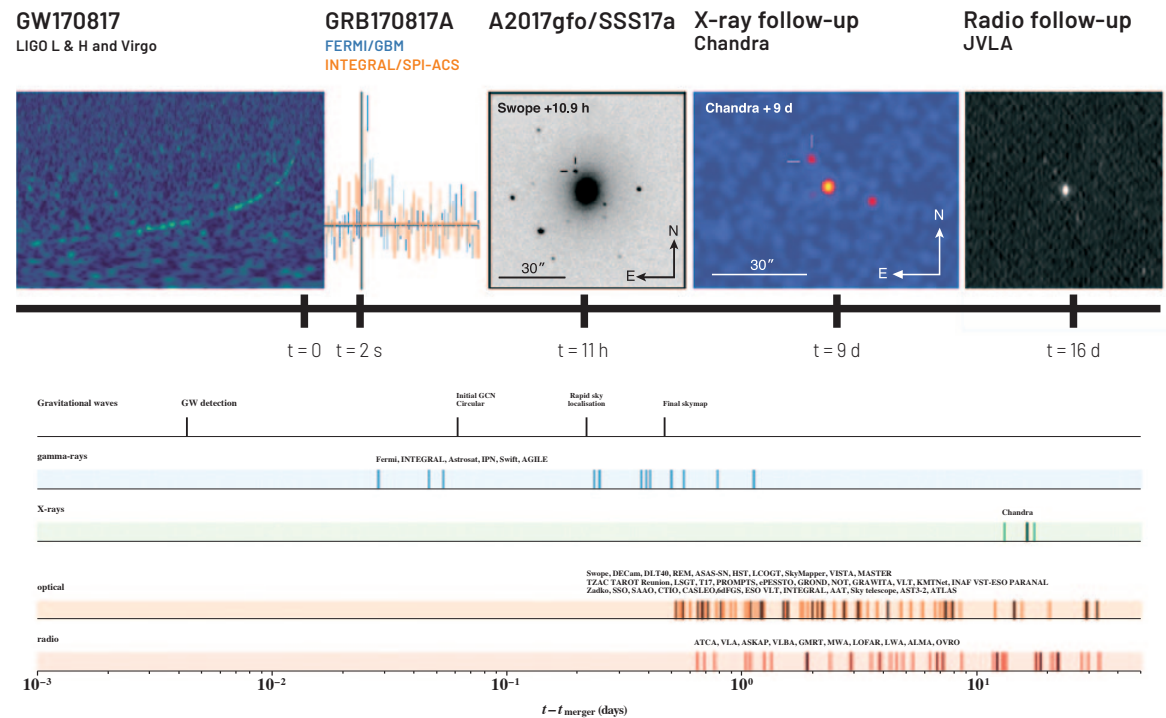

▲ *The working of multi-messenger astronomy: as one facility first discovers an event (left, in this case an outburst of gravitational waves from a merging binary neutron star), others may find it independently (second from left, gamma rays) or, alerted by the initial discovery, search and find it at other wavelengths (optical, X rays, radio, from left to right). This vastly improves our ability to understand the event.*

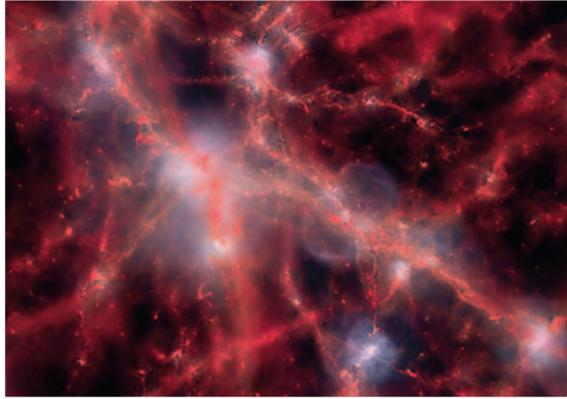

▲ *A rendering of the cosmic web using the EAGLE simulation, a very large effort to model the formation and evolution of the cosmic web and galaxies within it using supercomputer calculations.*

larger projects nationally coordinated. This is somewhat akin to what we have done for university participation in instrumentation three decades ago, but with some differences: software does not require a lab and is even more closely entangled with the work of most astronomers, so a distributed organisation makes more sense. Also, software efforts can have a wide range of scales, from being done by a single person to needing large consortia. Only the larger ones need national funding schemes and a coordinated effort, but all benefit from a system of proper teaching of software skills and a well set up system of software curation and documentation.

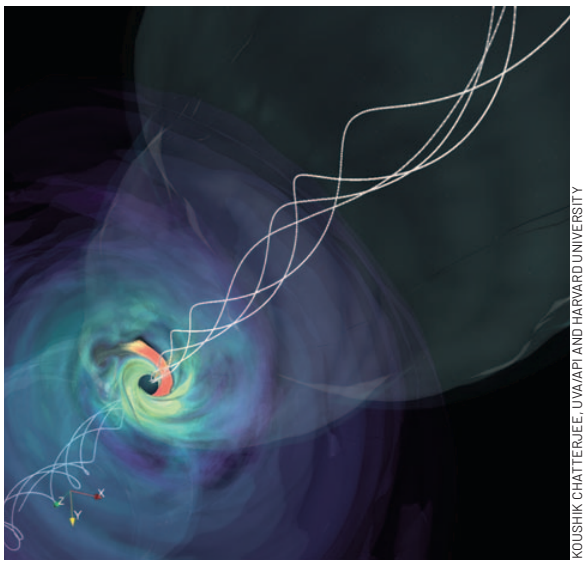

▲ *Visualization of the accretion of matter near the black hole at the centre of our galaxy, using the H-AMR code developed in Amsterdam. These simulations help understand the recent images of the black hole made using the Event Horizon Telescope.*

We primarily need agile top experts in our institutes, who are involved in frontline research and/or teaching in these domains. They will keep our community up to date and form active collaborations with adjacent fields, and they will find an attractive and challenging work environment in our community, inherent to our ambition level. By their interdisciplinary attitude, they will also make sure our efforts are outward-looking and synergize with those in adjacent fields. At the same time, our students will need a more advanced level of training in software and computing. This will not only increase their competitiveness in astronomy, but also in the many other professions they will be active in after graduation.

Also here, top quality comes from providing these experts and well-trained students with top-notch computing facilities, and in this respect the Netherlands would certainly benefit from bringing its effort to a higher level, commensurate with our partners and competitors. This will require a mix of good national facilities (in concert with other compute-intensive disciplines) and aligning with other countries to get our community access to the top international computing centres. With an initiative in this direction, we can broaden the leadership position we now have in some highly specific and somewhat isolated areas to include a large fraction of our scientific effort.

Astronomy has also been one of the pioneer disciplines in open science. As one of the earliest disciplines publishing via arXiv.org since its founding in 1991, rapid and open sharing of publications has become second nature to us. We have a similarly long tradition of open sharing of data, as a natural by-product of shared international use of publicly funded observatories and their need to keep good data archives. Our ambitions in joint software production and collaborative data science and compute-intensive research will naturally lead us into further sharing, using the FAIR (Findable, Accessible, Interoperable and Reusable) and open-source concepts in further support of reproducible research and optimal use of resources.

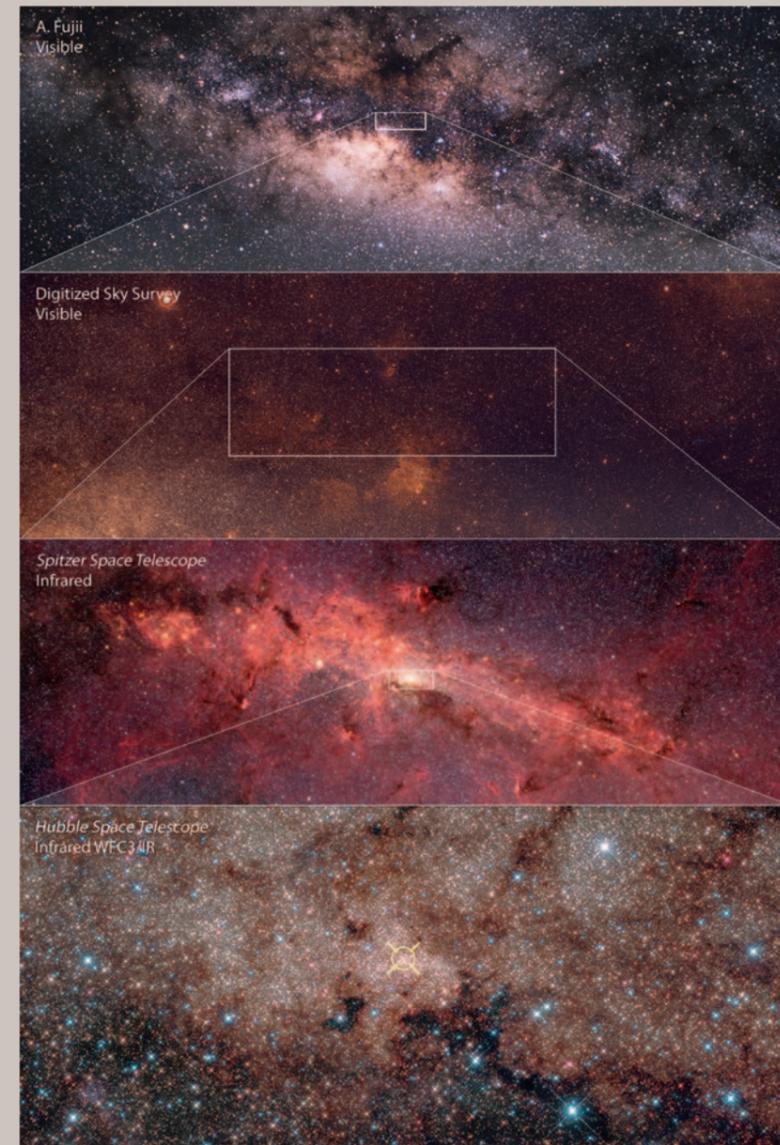

◀ *Zooming in to the centre of our Milky Way galaxy: The top image shows the central bulge, over 10.000 light years across, in starlight with dark absorbing dust clouds in front. The zoom in shows a region 2000 light years across near the centre, with most starlight hidden behind dust. The next zoom is 1000 light years across, in infrared light with Spitzer that can penetrate most dust clouds, showing the star behind them. The bottom panel shows the centre-most 50 light years imaged with Hubble in infrared light, with many stars near the centre and the actual location of the centre of the Milky Way (Sgr A\*) marked with the yellow circle.*

## 2 ASTRONOMY IN CONTEXT

The Universe is a source of wonderment and inspiration. Astronomy invites us to think about our cultural identity, raison d'être, and place in the cosmos. Children almost invariably ask questions on topics ranging from the Moon to inhabited worlds other than ours and from space travel to black holes. Often this curiosity persists throughout life. Astronomy's interaction with society is therefore very broad and includes engagement with the general public, education at all levels, human capital development, technology transfer, and assessing our own internal culture and impact. In this chapter we discuss a somewhat diverse set of topics in this category and assess what action is needed from us.

## 2.1 DUTCH ASTRONOMY - A REPUTATION FOR EXCELLENCE

Astronomy in the Netherlands has a track record of excellence that started at the end of the 19th century following a change to the educational system[1]. The generations of pioneering world-class researchers who as a result shaped 20th century Dutch astronomy include many names familiar to today's astronomy students: Jacobus Kapteyn, Willem de Sitter, Anton Pannekoek, Jan Oort, Marcel Minnaert, Henk van de Hulst, Adriaan Blaauw, Kees de Jager. This line of excellence continues unabated. Over the last 25 years seven Dutch astronomers have won the Spinoza Prize, the highest academic distinction in the Netherlands. The Kavli prize ('the Nobel Prize for astrophysics') was awarded to Ewine van Dishoeck in 2018, while the detection of the first gravitational waves and the first image of a black hole earned the teams, including a number of Dutch astronomers, the Breakthrough Prize ('the Oscars of Science') in 2016 and 2019, respectively; Heino Falcke was awarded the 2021 Henry Draper Medal. The Netherlands is well known abroad as the source of 'tulips and astronomers for export', as many Dutch astronomers pursue a career outside the Netherlands.

A key indicator of the success of Dutch astronomy is the high number of prestigious American postdoctoral fellowships awarded to Dutch PhDs (see figure below). Dutch PhDs received the largest number of prize fellowships after the US, and per capita the Netherlands is even ranked first worldwide. A second key indicator of success is the large number of Dutch astronomers who received an Advanced Grant from the European Research Council (16 over the period 2008-2019, see figure bottom). Dutch astronomy has received the largest number of grants per capita over this period. In both these cases, the Netherlands can be seen to be leading in the field and this underlines the high level and the visibility of the Dutch astronomy programme. Other success indicators include the often-leading role of Dutch astronomers in European and global organisations and advisory committees.

The quality of Dutch astronomy is regularly reviewed by independent international visiting committees. In 2010 NOVA was evaluated as 'exemplary' in a review from the OCW/NWO, and most recently this status was reconfirmed by the NOVA International SEP Review in 2016. In its recent 2017 review, conducted by an international panel of experts on behalf of NWO, ASTRON received the highest rating of excellent. A similar review of SRON in 2018 concluded its strong international reputation - 'excellent in space research' - is far beyond what might be expected from an institute in a small country such as the Netherlands. A 1999-2009 impact study by Times Higher Education among 138 countries in astronomy and astrophysics[2] ranked Dutch astronomy in sixth position, i.e., firmly belonging to the group of traditionally strong astronomical countries such as the USA (#5) and the UK (#8). A more recent impact study by the Scimago Journal and Country Ranking[3] over a longer period (1996-2019) concluded that the Netherlands maintained its top position just below the USA, and ahead of the UK and Australia. We believe that this level of international impact for a country as small as The Netherlands is possible precisely due to the national strategic coordination that we have implemented now for over half a century.

The impact of our work and training extends much beyond our own field, however, since we train our students more broadly. Our MSc, PhD, and postdoc programmes deliver graduates with sharp analytical and solution-oriented skills and with expertise in methodologies that play an increasingly important role in our society, such as visualisation, data mining, Bayesian inference and Artificial Intelligence. These students have broad interests and find their way into many sectors outside astronomical research as well. Many young talents that are attracted to our country by our programmes stay afterwards for jobs in high-tech fields. Our (hardware and software) technology similarly has a significant impact on society, as detailed in sects. 2.5 and 1.3.4.

## 2.2 INSPIRATION, EDUCATION, OUTREACH

All our institutes host outreach and education departments, with expert professional staff in this field. These experts help and support us to engage with the public at large and develop and run educational programmes that support primary and secondary school curricula. Policy makers and industry view astronomy as an important ambassador of STEM disciplines at all levels of education, essential for the future prosperity of our society.

Our overarching goals for the next decade are:
- Continue to inform the general public about

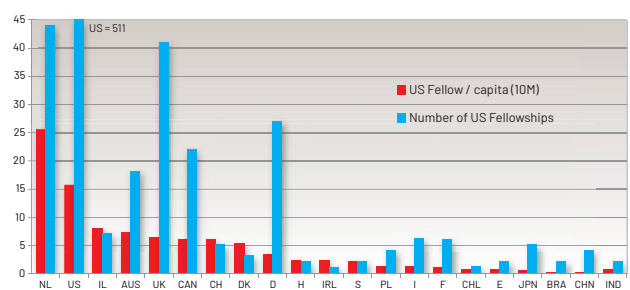

**S FELLOWSHIPS (-2020)**

▲ *The total number of highly competitive American fellowships awarded by NASA to PhD graduates of different countries for the period 2008-2020 (blue), and per capita (red). The Netherlands received the largest number of Fellowships after the US. Dutch PhDs received the largest number of fellowships per capita. (Source: NSF).*

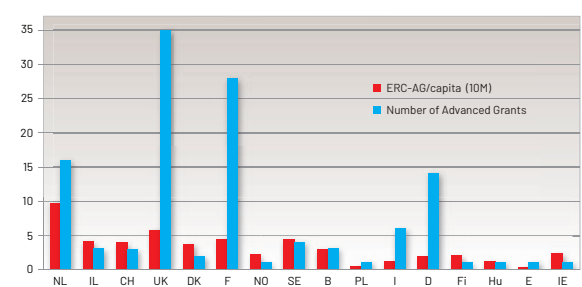

**ERC-AG ASTRONOMY GRANTS (2008-2019) PER 10 MILLION INHABITANTS**

▲ *The number of ERC Advanced Grants in astronomy (PE9) per country in 2008-2019 (blue), and normalised per capita (red). Dutch astronomers obtained 16 grants over this period, ranked 3rd after the UK and France, and ranked 1st in grants per capita.*

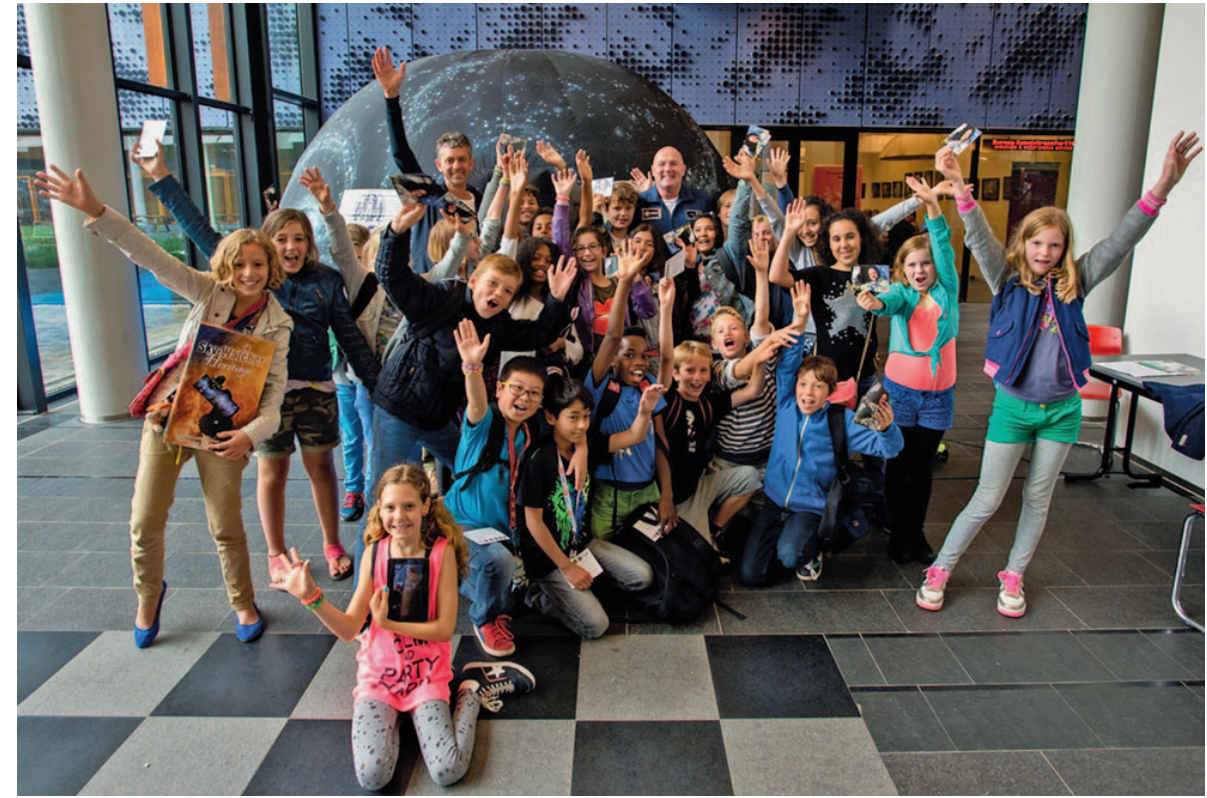

▲ *Bringing the skies to school children everywhere using the NOVA mobile planetarium.*

[1] *The introduction of the HBS, a new type of secondary school with a modern, science-oriented curriculum, as well as some reform of the universities.*

[2] *https://www.timeshighereducation.com/news/top-countries-in-space-sciences/408577.article*
[3] *https://www.scimagojr.com/countryrank.php*

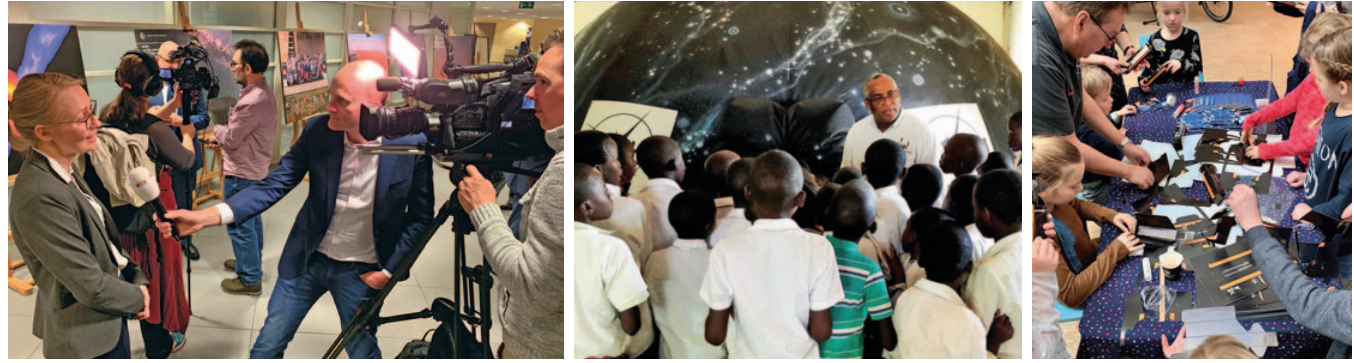

▲ *Monika Moscibrodska is interviewed by NOS about the first image of the M87 black hole with the Event Horizon Telescope. Middle: Outreach programme in Namibia associated with the scientific collaboration on the African Millimetre Telescope (left). (European press conference, Brussels, 19 April 2019). Right: Primary school students building and practicing with the NOVA developed DIY spectroscopy kit. (International Day of Women and Girls in Science in February 2019 in Leiden).*

exciting and inspiring discoveries
- Promote trust in the process and values of science and exploring ways in which astronomy can help tackle societal challenges
- Innovate in education and public engagement by expanding from (one-way) classical science communication to true (two-way) public engagement and dialogue initiatives using modern techniques and media channels
- Further increase the reach of 'In Classroom' education, with special attention for groups from underprivileged backgrounds and with low science capital, to foster equity and inclusion
- Continue participation in Ministry of Education, Culture and Science commissioned curriculum and exam programme evaluations
- Train the next generation of astronomers to become engaging communicators, contributing to diversification of career paths
- Expand activities to the Caribbean parts of the Kingdom and strengthen collaborations with partner countries

An important part of our societal responsibility is teaching astronomy at all levels, from elementary astronomy to advising MSc thesis projects. SRON and ASTRON staff contribute to these programmes, teaching specialised courses or supervising projects jointly with NOVA staff. After their MSc, roughly half of our graduates enroll in a PhD programme after their MSc and roughly half leave academia and pursue careers elsewhere. We therefore recognise the importance of preparing students for a diverse range of career paths. We organise not only local activities for this, but also an annual national career event for astronomy MSc students. The programme includes presentations and career advice from astronomy alumni on various career paths, as well as networking opportunities. The Netherlands is also coordinator of the EU-funded SKIES[4] project, that is specifically aimed at developing entrepreneurial competences of astronomy PhD students and postdocs.

## 2.3 OUR COMMUNITY – EQUITY, INCLUSION, DIVERSITY

In many fields in our society there are underrepresented groups: e.g., boards of big companies, professorships at universities, and in our own national astronomy community at every level. Across the board, we can improve diversity, equity, inclusion, and accessibility in Dutch astronomy and in the past few years we have begun to critically examine our own community and ask ourselves how we can improve in these aspects. We recognise that the systematic barriers faced by different underrepresented groups can be diverse and require a nuanced and intersectional approach.

Fostering an inclusive and equitable astronomy community will lead to a more diverse and representative workforce, which is a key component in increasing scientific excellence and creativity. This work is challenging and requires a multi-pronged approach, with sustained commitment. Our astronomical institutes are already collaborative by nature and geographically close. Leveraging on this, and to learn from each other's shared experiences, in 2020 we formally combined our individual institutes' efforts nationally in the NAEIC (National Astronomy Equity and Inclusion Committee). Through NAEIC[5] we aim to represent and support the 900+ astronomers, instrumentalists, engineers, support staff, and students who work in Dutch astronomy.

The NAEIC comprises representatives from the RvdA institutes together with an independent chair and observer from the RvdA; it is an advisory committee that reports to the RvdA. After establishing our mission statement and a universal Code of Conduct adopted by all Dutch astronomy institutes, we have undertaken several high-priority activities. These included training as point of contact advisors in the community, masterclasses for our committee members and institute management and work with ECHO[6]. We also organise live-captioned monthly webinars focused on equity and inclusion, a "creating your own place in astronomy" programme for PhD students and postdocs and equity and inclusion sessions at our national NOVA PhD schools. We have recently hired an equity and inclusion officer, to add expertise we did not have. This officer will lead development of a strategy for addressing how to support and create an inclusive astronomical community on the national scale.

Many of our members are also involved in global efforts, so we are working closely with similar committees and initiatives internationally. Examples are these include at the European Astronomical Society and the International Astronomical Union; we also aim to inform, and be informed

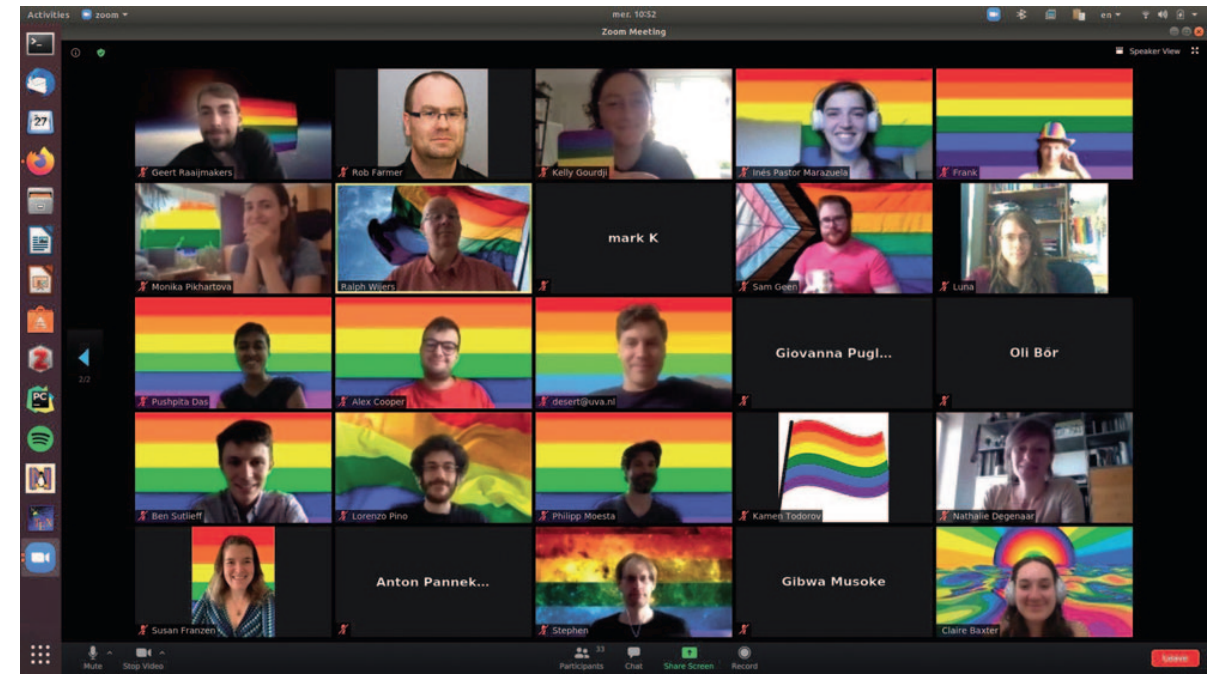

▲ *As NAEIC's early days overlapped with the COVID-19 pandemic and its lockdowns, we found new problems to consider and new ways of celebrating our diversity.*

by, their programmes, such as the 2021 IAU Springboard to Action[7], particularly in the context of global astronomy projects. This is a long-term, global effort that needs to target (potential) astronomers of all ages, from elementary school on.

Short term goals include (I) implementation of ECHO facilitated masterclasses in awareness of equity and inclusion for the entire Dutch community, including early career scientists and our support staff; (II) providing recommendations on inclusive hiring at different levels; (III) understanding how to improve on accessibility measures for all astronomers; (IV) learn how to provide a safe and healthy environment by eliminating harassment and bullying in our academic hierarchical structures. We are closely aligning these goals with our national science communication efforts, to ensure they reach those from disadvantaged areas and backgrounds and are mindful of collaboration opportunities with other Dutch efforts. A prime goal will be to work with our funding agency NWO to, e.g., assess and restructure funding processes so that they support more equitable, robust and adaptable resource distribution. This in turn would leave the community less vulnerable to exploitative power structures, as well as supporting those with significant responsibilities outside work (such as family or dependent care). This is especially important when such responsibilities become extra demanding at times, as illustrated by the recent COVID-19 pandemic.

---

[4] https://www.accessastronomy.eu/netherlands

[5] https://www.accessastronomy.eu/netherlands
[6] the Dutch diversity expertise centre: https://echo-net.nl/en/

[7] https://www.iau.org/news/pressreleases/detail/iau2101/

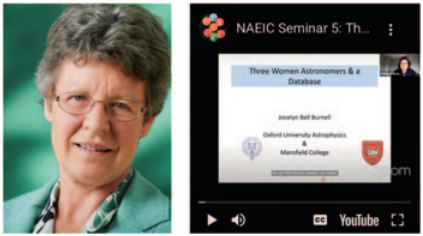

▲ *Professor Dame Jocelyn Bell honoured NAEIC by being one of the first speakers in our national EDI seminar series.*

Within the next ten years, our overall goal is to make equity and inclusion integral to our scientific community and our activities, and indeed that our community is representative of the diversity and breadth of talent in the wider modern Dutch society. We will work closely with primary and secondary education, and attempt the nation-wide implementation of local efforts such as Altair[8]. Moreover, we are fully cognisant that our work must continually adapt to societal change. Finally, we aim to expand the current paradigm of what a career in astronomy can encompass, to support these choices and to better include historically excluded groups.

## 2.4 ASTRONOMY AND CLIMATE CHANGE

Human activity, and in particular the burning of fossil fuels, has raised the atmospheric $CO_2$ concentration to levels that are unprecedented in human history. There is no scientific doubt that humans have been the main cause of the observed global increase in temperature for about 50 years now. More and more measures against pollution and $CO_2$ emission are being taken worldwide and there is only a limited amount of $CO_2$ that can still be emitted if we are to prevent warming above 1.5-2 degrees that the IPCC considers dangerous. Several studies indicate that astronomers worldwide contribute quite significantly to emissions due to travel and computing. At the same time, the astronomical perspective on Earth as a tiny planet in the vast Universe generates unique opportunities to contribute to the discussion and awareness about the climate crisis.

Our community can play a part in addressing the climate crisis in at least four areas:
- Quantify our field's current impact on climate and in particular $CO_2$ (computing, travel, scientific operations etc) and devise and implement measures for reducing it.
- Quantify the impact on climate of our future plans (e.g., from new infrastructures) and develop ways to minimize their environmental impact with minimal impact on scientific excellence.
- Contribute to climate/sustainability knowledge, for example with data (e.g., long-term astro-climate data of remote sites that are usually not covered densely by weather stations) or the exchange of knowledge/expertise (e.g. on exoplanet climate) with professional climate researchers.
- Building awareness. Astronomers can offer a specific perspective: the special place of the Earth in the Universe (cf. the "Earth rise" and "Pale Blue Dot" pictures), in combination with spectacular outreach opportunities that inspire the public.

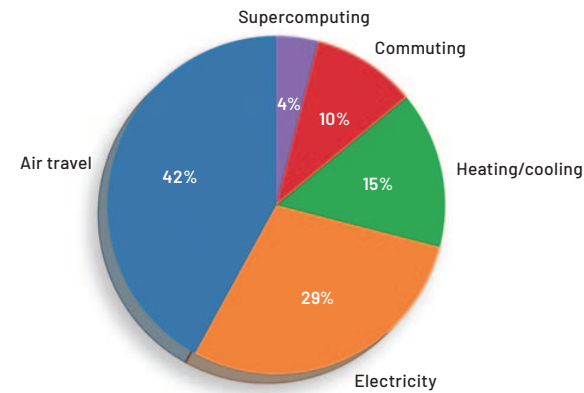

▲ *Result from our study into the relative contributions of identifiable contribution to our carbon footprint in 2019.*

Many climate-change initiatives have been started in the last few years throughout the astronomy community: in large collaborations (Athena X-IFU, LIGO/Virgo), in observatories/departments/institutes (e.g., ESO, CFHT, MPIA), countries (US, Canada, Australia), professional societies (EAS, IAU) and special groups (Astronomers for Planet Earth) as well as funding agencies such as NWO. We have formed an RvdA working group on sustainability (including climate change) that will bring together different people inside Dutch astronomy and reach out to other disciplines and advise us.

As a first project, we made a 2019, pre-COVID, $CO_2$ emission benchmark that shows that Dutch astronomy emits nearly 5000 tons of $CO_2$ per year, excluding emissions that could not yet be tracked[9] (e.g., from large collaborations, observatories,

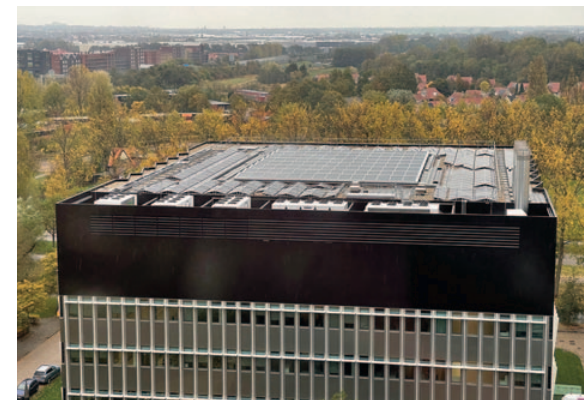

▲ *As our infrastructure is modernized, ensuring is has a naturally lower carbon footprint is a high priority. Shown are solar panels on the roof of the new SRON building in Leiden.*

space agencies and part of international supercomputing). The RvdA is implementing a $CO_2$ policy aiming at significant $CO_2$ reduction by 2025, by an amount that each of the participating institutes will determine in close consultation with their staff to secure support for the measures. The sustainability committee and the institutes will monitor the progress and the committee will expand its activities to other aspects of sustainability. There are of course delicate balances to be kept with the needs of international community building and networking, and equitable access to those for all researchers. This is an active area of discussion, involving collaboration between the sustainability working group and the NAEIC.

## 2.5 ASTRONOMY AND TECHNOLOGY

Astronomical research is a powerful driver of technological innovation. Extracting as much information as possible from the weak signals that have travelled through the Universe for billions of years to reach us poses extreme challenges. The instruments with which we capture these signals represent the latest and finest that human technology and ingenuity can achieve. The constant innovation required in instrumentation to stay at the forefront of astronomical research benefits other areas of society too. We have numerous examples of successful technology spin-offs from astronomy, from the CCD detectors in phones and digital cameras, to infrared thermometers and WIFI, to name just a few.

The Netherlands has a strong track record in delivering world-leading technologies, instruments, and infrastructures for the astronomical community thanks to the coordinated instrumentation programmes of ASTRON, SRON, and NOVA. Each of these entities represents the national base for Dutch contributions and research support to the international programmes of the SKA, ESA, and ESO, respectively. These technology and instrumentation programmes are an integral part of our long-term science strategy, and all three organizations develop and maintain core technical expert-ise and facilities. Those core capabilities ensure we are preferred partners capable of making unique and leading contributions to ESO, ESA, and the SKA. Those in turn help secure leadership roles for Dutch scientists in the large international consortia for new instruments or facilities. These consortia then define the scientific programmes and provide us with early access to data and guaranteed observing time once operational.

[8] https://api.uva.nl/content/news/2020/12/nwo-diversity-initiative.html?cb

[9] *Nature Astronomy* 5, 1195-1198 (2021); https://www.nature.com/articles/s41550-021-01552-4

Our technology development at NOVA, ASTRON, and SRON is also a natural nexus point for our partnership with the technical universities, other knowledge institutes in the Netherlands, and industry. Partnerships with technical universities bring collaboration opportunities on high-tech, low technical readiness level (TRL) research relevant to astronomical instrumentation. At the same time, a strong collaboration with the commercial high-tech industry sector in the Netherlands and abroad creates both spin-in, where astronomy makes use of new technologies developed for commercial applications, and spin-off, where innovations made for scientific research find applications in commercial products. These industrial collaborations represent an important valorisation channel for national investments in Dutch astronomy. Recent examples are the partnership between Airbus and SRON to deliver the SPEXone instrument onboard NASA's PACE mission, the companies participating – with ASTRON – in the design and construction of both LOFAR and the SKA, or the 26 Dutch Small/Medium companies (SME/MKB) that contributed to the NOVA development of the MIRI instrument now flying onboard JWST.

The science programmes of ESO, ESA, and the SKA extend over long periods, typically 10-30 years into the future. NOVA, SRON, and ASTRON therefore maintain long-term roadmaps for potential necessary future instrumentation and the technology development required to bring these instruments and facilities to fruition. It immediately follows that these developments require dedicated and sustained investments extending over periods far beyond any single grant or even decadal planning cycle. We discuss these technology roadmaps further in this section.

### Radio Astronomy (ASTRON)

We are exploring redevelopment of the Westerbork site as one of the options for future investment in order to ensure that the Netherlands maintains a technology development base for SKA-mid type systems; at the same time this gives us an operational VLBI platform as part of the EVN for the next 20 years and a link to the SKA-VLBI. We have developed significant big-data expertise and infrastructure at ASTRON for LOFAR and SKA.

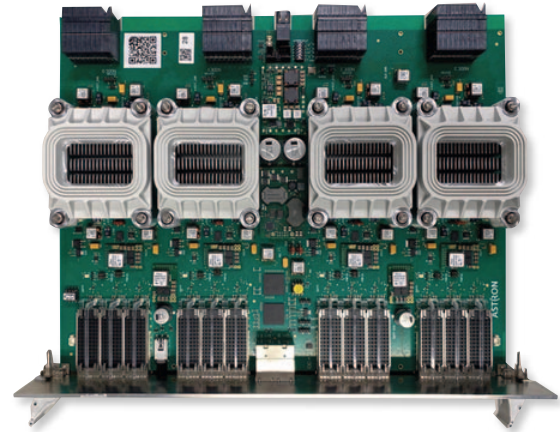

▲ *The UniBoard 2 processing unit shown here is an essential part of the LOFAR2.0 upgrade. It will allow LOFAR stations to generate more station beams to facilitate more simultaneous observations and in two frequency bands. Incoming radio signals are filtered, beamformed, and correlated all on this board, increasing the data rate to the LOFAR central processor to 9 Gbps (image credit: ASTRON).*

This allows us to forge ambitious collaborations addressing the increasing needs of big-data storage and processing in Dutch astronomy as well as with other disciplines. Examples are prototyping easy-access, adaptive big-data pipelines and production archives that can be used to address the combined future data needs of ASTRON, SRON, NOVA and the wider scientific community. In the coming decade ASTRON will continue its technology development programme in radio astronomy. We will develop integrated frontends to eliminate complex interfaces between analogue, digital and opto-electronic parts of current front-end designs, to achieve stable operations and affordable systems for next-generation radio telescopes. These front-ends are crucial to ensuring that the large-N interferometric arrays of the future can achieve high sensitivity, a large field-of-view and excellent dynamic range.

Long-baseline radio interferometry allows us to zoom in on the inner regions around black holes and young stars in formation. This depends on global networks of radio antennas and thus the need to include also those located in challenging RFI environments. Advanced RFI mitigation techniques are needed to allow operation of sensitive radio telescopes in these increasingly noisy environments.

The ever-increasing complexity of radio astronomical data sets and their calibration is beyond the expertise of most scientific users. It is therefore critical to distil these data sets to science-ready data products that can be used by a broad range of astronomers. To achieve this, we will develop newly designed, integrated hardware/software systems; they incorporate prior knowledge of instrument performance and more advanced modelling of ionospheric effects and thereby result in smart processing of the large raw data streams into science-ready products.

Data volumes from modern radio telescopes, like LOFAR and SKA, are extremely challenging. Science teams need to be able to bring their novel algorithms and analysis to the data without having to invest in major in-house computing infrastructure. The aim of ASTRON's Science Data Centre initiative is to facilitate "big data" research while also complying with modern standards requiring scientific data to be Findable, Accessible, Interoperable and Reusable (FAIR), in line with our broader ambitions in software development (sect. 1.3).

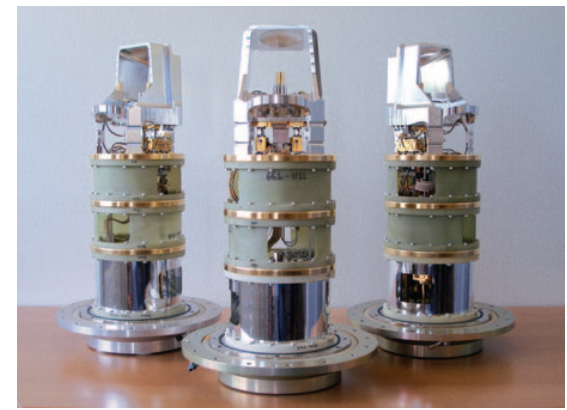

▲ *Receivers for the ALMA telescope, made by NOVA based on technology partly developed at SRON.*

### Optical-infrared and sub-mm (NOVA)

The first generation of instruments for the Extremely Large Telescope represents a major leap in the size, cost, complexity, and tolerance of optical-infrared instruments. To cope with these demands, NOVA is investing in a new manufacturing facility with a 5-axis milling machine that can handle parts with dimensions of 1,5 m, and machine them with micrometre-accuracy. We will install at the same facility a matching coordinate measurement machine that can verify the dimensions of manufactured parts, under clean conditions. We are realising this facility in collaboration with industry, and it is located at a company that specialises in metal fabrication. It will be a centre for knowledge exchange and education in high-precision manufacturing in the Northern Netherlands.

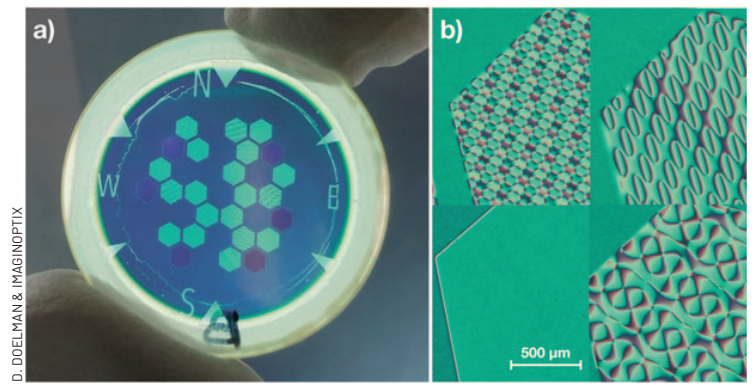

▲ *Using liquid crystal micropatterning, a Holographic Aperture Mask has been produced. It combines high light collecting efficiency with high resolution imaging, and at the same time some spectroscopic capability (R~15). It was successfully tested in 2019 using the OSIRIS imager at the Keck telescope.*

Over the coming decade we will expand NOVA's network of industry partners beyond the current focus on precision manufacturing and mechatronics, to micro-photonics and additive manufacturing. This enhancement will be of special importance for the EPICS instrument, the high-contrast imaging and spectrograph for the ELT we are currently developing. In NOVA we have built up a strong technology position in coronagraph design, polarimetry, AO control and systems engineering vital for realizing this instrument. Continuing this development and demonstrating the key technologies in a testbed or pathfinder instrument is a key priority of our NOVA technology programme.

In sub-mm instrumentation, we currently focus on significantly increasing the RF/IF bandwidth while simultaneously reducing noise levels. These developments are based on the tight integration of low noise HEMT (high electron mobility transistor) IF and low frequency RF HEMT amplifiers. They are part of the current ALMA Band 2 project but will also enable upgrades of other bands such as Bands 9 and 6. Parallel developments on sideband separating receivers, and on digital front end technology, will further improve the performance and robustness of the receivers. The next major step in astronomical sub-mm receiver development will be multi-pixel receivers, directly multiplying the science output from the telescope. We will do the necessary research towards reducing heat dissipation in the amplifiers, and to approaching the quantum sensitivity limit at 100 GHz - 1 THz frequencies. Separately, we continue work towards developing large-format Microwave Kinetic Inductance Detector arrays (MKIDS) for a future single-dish sub-mm observatory.

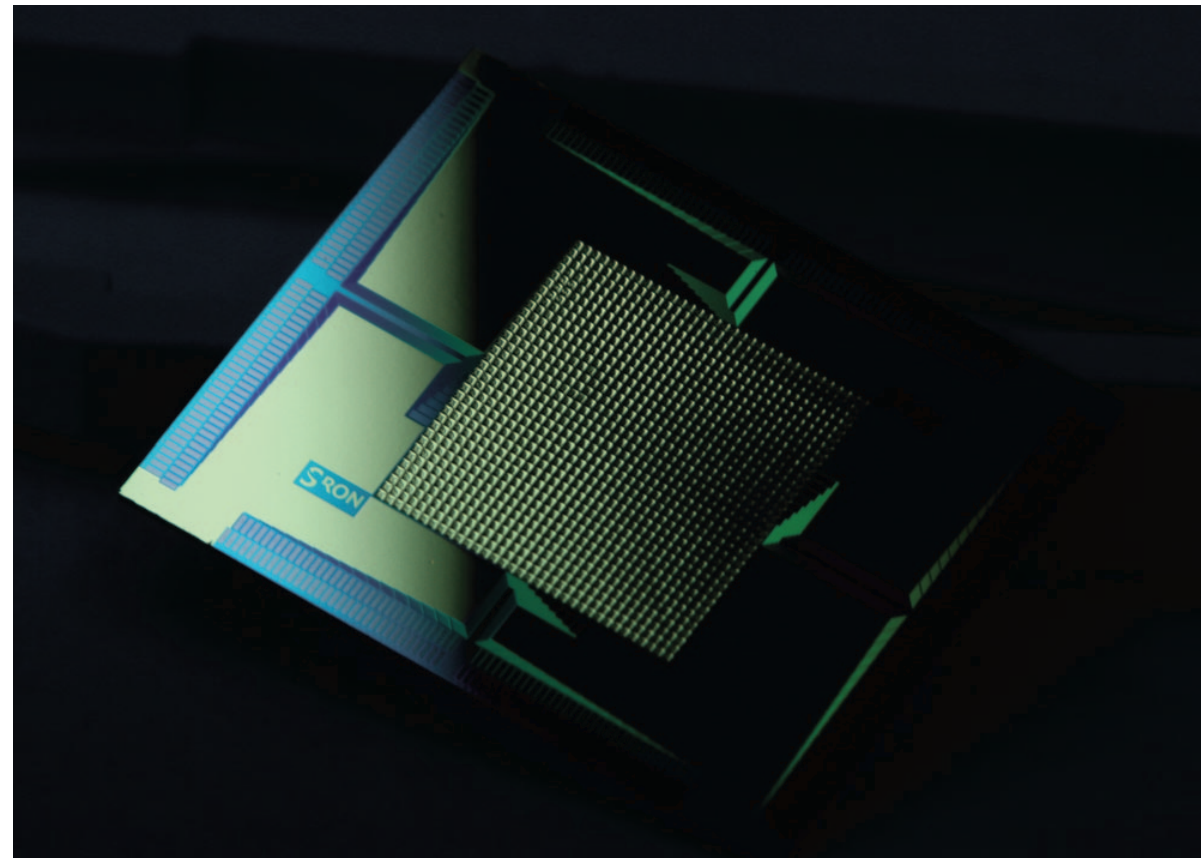

▲ *This image shows a 32x32 pixel X-ray TES (Transition Edge Sensor) array, uniformly fabricated on a 19 mm x 15 mm silicon chip. The pixels of the detector are 240 µm x 240 µm and closely arranged with a pixel gap of 10 µm. TES arrays like these will power the next generation of ultra-high spectral resolution instruments on upcoming X-ray missions such as XRISM and Athena.*

### Space Instrumentation (SRON)

The coming decade SRON will mainly focus on instrumentation and R&D for missions in the X-ray and far-infrared, as well as the gravitational wave interferometer LISA, building on a long history of work in the relevant technologies. Smaller-scale efforts in the exoplanet missions PLATO and ARIEL, as well as scientific support (instrument calibration for various missions, and continued development of software analysis tools such as SPEX) are also underway.

ESA's L-class X-ray mission Athena, to be launched in 2034, will include the X-ray Integral Field Unit (X-IFU) instrument that will provide us with unprecedented spatially resolved, high-resolution X-ray spectroscopy. SRON is the co-PI institute, responsible for the detector assembly including the TES sensor and electronics, using an NWO Roadmap grant. We are already investigating ways of further improving the TES detectors for future, even more advanced missions. We focus on achieving larger format arrays and accompanying readout systems, anticipating larger fields of view while maintaining sensitivity and very high spectral resolution. For the even further future, we are investigating the feasibility of X-ray Interferometry, a means to achieve ultra-high-resolution imaging of distant objects, which ESA has identified as an important technology route in the recently released Voyage 2050 report.

The Netherlands will take a similarly leading role in the development of the LISA gravitational wave observatory through a joint development programme between SRON, Nikhef, and TNO to deliver low-noise quadrant photodiode detectors, flight electronics, and optical-bench mechanisms. We have recently submitted a Roadmap proposal to fund gravitational wave research, including these LISA contributions.

SRON also has a long-standing track record in cryogenic, heterodyne detector systems for far-infrared (FIR) and sub-mm astronomy. These systems are all very mature and have flown on previous space missions including the HIFI instrument onboard the Herschel Space Observatory. Several proposed future FIR missions have such instruments as their baseline design. In addition, we are also developing new technology for FIR astronomy: ultra-sensitive MKIDs arrays, which provide excellent sensitivity and superior multiplexing properties. This makes them well suited for use as large-format imagers, spectrometers employing optical front ends, or arrays of high-resolution, on-chip spectrometers. The performance of our current MKIDS detector systems is world-leading and they are the preferred technology for future NASA and ESA FIR missions. Based on this technology, we are actively working with multiple US-led teams to develop FIR mission concepts with a proposed launch date in the early 2030s.

Finally, compared to data sets at other wavelengths such as radio, the volume of data from our space missions tends to be much more modest (e.g., due to satellite telemetry limits). The high complexity of the data, however, makes the

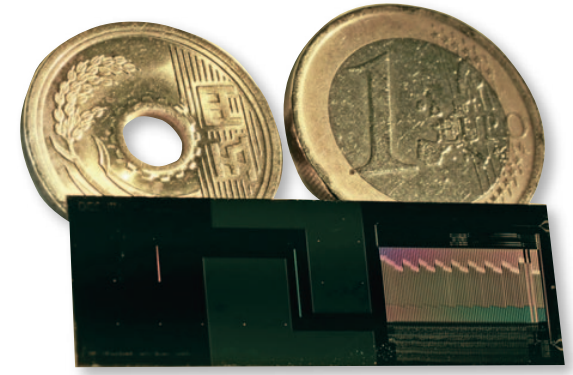

▲ *Ultra-sensitive MKIDs arrays represent a powerful new technology for FIR astronomy. They provide excellent sensitivity and superior multiplexing properties making them well-suited for large-format imagers and high-resolution, on-chip spectrometers. The DESHIMA-1 MKIDS-based on-chip spectrometer shown here contains 49 channels and was successfully demonstrated on Atacama Submillimetre Telescope Experiment (ASTE) in 2019. An upgraded DESHIMA-2 (350 channels) is planned for late 2022.*

modelling and analysis necessary to interpret them computationally expensive. We have over 50 years of experience developing tools for both high-energy and FIR spectroscopy and currently maintain the SPEX and RADEX analysis packages. In the coming decade, we will build on this effort by upgrading the existing tools to a more open and accessible software ecosystem and developing new analysis techniques and tools that take advantage of modern techniques in data science. We will match these upgrades with improvements in the underlying atomic data necessary to model and interpret observations with these new instruments. Like their hardware analogues, we can use these technologies as possible Dutch contributions to future missions; an example of this is our participation in the THESEUS gamma-ray burst mission recently proposed to ESA. In any case, these developments represent an important asset for the Dutch astronomy community during the science harvesting phase. They will be relevant for missions including XRISM, Athena, ALMA, and JWST, and serve as seed examples for our planned more structured approach to software development and curation (sect. 1.3).

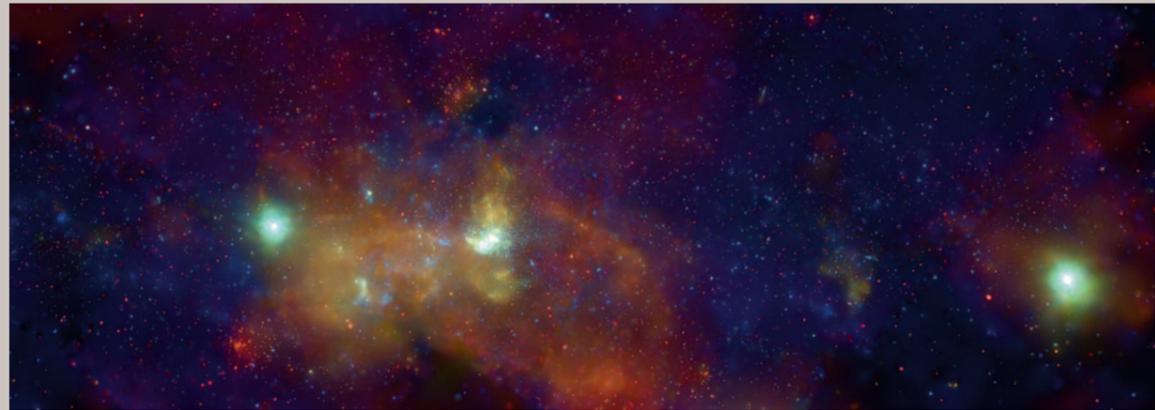

▼ *The central 1000 light years of the Milky Way imaged in X rays using the Chandra satellite. The central source is again Sgr A\*, and different colours indicate different shapes of the X-ray spectrum of the sources.*

# 3  THE ROAD AHEAD

To map our road ahead we have to make some assumptions about our environment and the state of science funding in the coming decade. We adopt the somewhat conservative approach of assuming that we can continue to sustain the level of activity that we have successfully kept in the past few decades (see sect.4.1). We assume constant levels of permanent staff, excepting the enhanced effort in data science and computing, and the same level of availability of grant funding and successful proposals. We also demonstrate that we can realise our ambitions with a constant funding envelope for all of Dutch astronomy (in 2020 Euros). Here we summarize the main priorities that drive our strategy for the coming decade (3.1) and give some details of the funding sources we aim for and total expected expenditure (3.2).

## 3.1  PRIORITY NEW INITIATIVES FOR THE DECADE

From the lines of argument in the previous two chapters, we have derived our funding priorities for the coming decade. They follow the main principle that top astronomy is top talent plus top facilities and are designed to keep our community strong, connected, and positioned at the forefront of astronomy research. They also build on decades- long lines of technical and industrial expertise that give us an edge in certain instrumentation areas. With each funding priority we address multiple science questions and for each we require multiple astronomy institutes in the Netherlands: collaboration across the Dutch astronomy landscape is essential for them to be realised. The priorities are

1. Structural funding for NOVA, directed at:
   a) maintaining and enhancing optical-infrared instrument capability
   b) high impact in data- and compute-intensive astronomy
2. Investments in three major facility areas:
   a) a leading role in next-generation ELT instrumentation
   b) a critical contribution to the LISA gravitational-wave space mission
   c) a frontier role in radio astronomy by delivering and developing future technologies for SKA and LOFAR

Importantly, we note that these investments can only be fruitful in the context of a vibrant and healthy scientific community, with sufficient support for smaller projects and local infrastructure. We must coordinate the major investments strategically on the (inter)national scale, but a healthy scientific community has a mix of small and large research projects at any time. Such small projects can be small in total size, by one or a few PIs. They can also aim at getting us a small share in a large facility that is led by others, leveraging the facilities we lead as an asset through which we gain access to more top facilities than our community can support on its own (see, e.g., sect. 1.2.3).

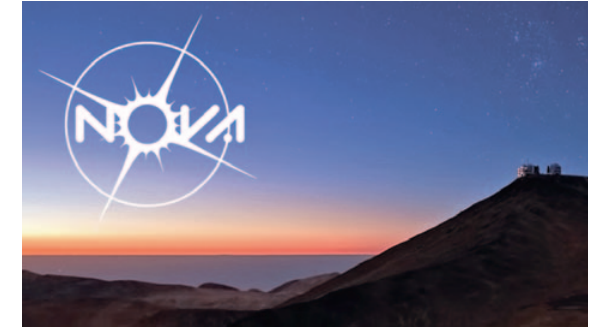

▲ *Long-term secure funding for NOVA will enable us to keep making leading contributions to major international facilities, with their very long time scales.*

### 3.1.1  STRUCTURAL FUNDING FOR NOVA

**Instrumentation and science programme**
Since 1998 Dutch astronomy has benefited greatly from targeted funding through the Dieptestrategie. We have used it to reinvigorate Dutch astronomical instrumentation work, with emphasis on ESO facilities (instrument efforts for space and radio facilities being primarily in the hands of SRON and ASTRON). As a result, NOVA is now an important partner in the ESO instrumentation programme. We thereby have a key avenue to obtain preferred access to the world class ESO telescopes, with instruments that are in tune with our specific scientific interests. Thus, through NOVA, we have enabled the university departments together to exploit the Dutch ESO membership to great advantage, by making strategic choices about which instruments to invest in, and in what role.

Long-term funding of NOVA is essential if we are to continue this success. As instrumentation for the large telescopes grows in scope and complexity, the trend is inevitably towards fewer, more expensive projects with longer schedules and larger technical and scientific teams. Continued consolidation at the national level is the only way for the Netherlands to remain relevant. With project timelines on the order of a decade, we as project partners must be able to provide guaranteed support (personnel and funding) over such periods.
It is therefore a priority for Dutch astronomy that NOVA be made permanent, with baseline funding secured over a sufficiently long timeline. With that, we can maintain the technical staff and expertise to enable participation in the next generation of large instrumentation projects. Over the past two and a half decades NOVA has been reviewed every five years and has consistently been evaluated as excellent/exemplary, with the review committees noting each time that the uncertain long-term future was the biggest issue.

Our scientific networks are an integral component of NOVA; they link astronomers at our different institutes into collaborative projects, and facilitate support and the exploitation of the data obtained with the facilities. NOVA has strategically funded PhD students, postdocs and permanent staff overlap positions to strengthen these networks' research activities, to make sure our instruments deliver the science, and to position Dutch astronomers for strong roles in international collaborations.

**Data- and compute-intensive astronomy**
Computing and data science are essential parts of many, if not most, aspects of astronomy. Methodology for these efforts, such as machine learning (e.g., transient searches in large data streams), data mining, advanced statistics and inference (e.g., neutron star parameters from NICER data), have become research areas in themselves. In the past we organised these efforts per project, with minimal cross-coordination or sharing of facilities and technical know-how. We now need better coordination, pooling of expertise and resources, and strategic investments, in recognition of the central role these activities will continue to play in shaping astronomy while projects continue to increase in size. To achieve the ambitions detailed in section 1.3, we will work together to put in place:

- High-level experts on permanent contracts in software (systems) development, data science, scientific computing, statistics, and inference to lead the effort, starting at approximately 3 fte per institute.
- A national programme of major software development and curation, of similar ambition level as our instrumentation effort, and with similar ongoing oversight and regular strategy exercises. It should set priorities, encourage wide cross-institutional collaboration on those, and consider all areas of software development.
- Better training in computing for all our students, and specialised graduation tracks for those who wish to emphasise computing and software in their future research or employment; the standards for this will be developed nationally.
- Development of standards of software development, documentation, and curation that improve the use and dissemination of all relevant software.

We thereby invest in 'professionalisation' of our computing and data science activities, which will allow us to make a big step forward in the effect and impact of our software and data science efforts. We also very much intend this as an outward-looking effort, linked to similar initiatives and investments in large-scale computing in other fields and in global astronomy, also expanding our role as pioneers in open science.

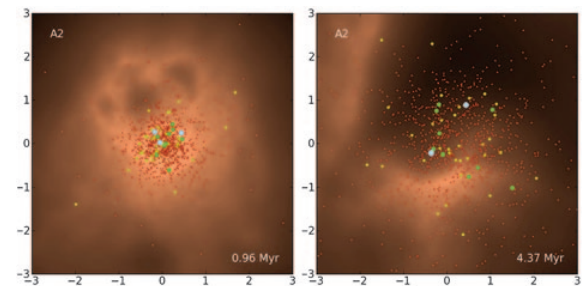

▲ Results from AMUSE, the Astrophysics MUlti-purpose Software Environment, developed to combine gas dynamics, stellar evolution, and stellar dynamics into a well-document framework. Shown is the evolution of a cluster of stars of varying mass (coloured dots) as they are born from a large molecular cloud (orange, brighter= denser gas) and start to heat and destroy that cloud.

### 3.1.2 INVESTMENTS IN THREE MAJOR FACILITIES

**Next-generation ELT instrumentation**
We have deemed active participation in ELT instrumentation a top priority for Dutch astronomy since our previous strategic plan. Securing long-term funding for NOVA will allow our national NOVA OIR laboratory to continue to play an important role in the VLT and ELT instrumentation programme. It will also allow us to do faster projects that can deliver a science return more quickly or can demonstrate a key technology for future use on the largest telescopes. For the ELT we specifically aim for:

**METIS completion and first science:** METIS is a general-purpose mid-IR instrument, one of the three first-light instruments for the telescope. The project is led from the Netherlands, with first science foreseen around 2028. It addresses some of our highest-priority science goals, such as direct imaging of rocky planets around the nearest solar type stars and the formation and evolution of distant galaxies.

**MICADO completion and first science:** MICADO is a near-infrared camera that will be the first instrument on the ELT, designed to produce images corrected for atmospheric blurring via active optics over a sizeable (arcmin) field of view. The Netherlands are senior partners (co-PI) in the project. Its science drivers address several of our main priorities: the dynamics of dense stellar systems, black holes in galaxies and the centre of the Milky Way, the star formation history of galaxies through resolved stellar populations, the formation and evolution of galaxies in the early Universe, exoplanets and planet formation, and the Solar System.

**R&D and design studies for future instruments – MOSAIC and EPICS:** the second generation of ELT instruments will not see first light until the 2030s, but we are already planning and scoping them. We need R&D and early involvement in building and designing these instruments to ensure a visible and scientifically attuned role. Two instruments are particularly interesting for the Nether- lands: MOSAIC, a multi-object spectrograph intended as a second-generation instrument for which NOVA could provide the visible channel; and EPICS, the high-contrast imager and spectrograph for exoplanet imaging and characterisation that is currently in a technical R&D phase.

Even with continued long-term NOVA financing assured, we will need investments from the national Roadmap for these projects.

**LISA: milli-Hertz gravitational waves from space**
Gravitational-wave astronomy promises to make major strides in the coming decades. Since the first, Nobel prize winning detection of gravitational waves in 2015, the ground-based LIGO and Virgo antennas have now identified more than 90 events, mostly mergers of black holes of 20-100 solar masses. Dutch astronomers and physicists are active in the science teams and in the planning of further upgrades that will push towards lower-mass black holes and neutron stars, and of new ground-based experiments (such as the Einstein Telescope).
Complementary to this, ESA has selected the Laser Interferometer Space Antenna (LISA) for launch in the mid-2030's. With LISA, we will for the first time explore the milli-Hertz frequency domain, in which we expect to find a rich variety of sources different to those seen from the ground. They offer fundamentally new science from a rich variety of sources that are not accessible from the ground. It can 'see' supermassive black holes at the dawn of the Universe, complementing our view of the earliest galaxy formation. We can also use it to measure more nearby ones so accurately that General Relativity will be tested to unprecedented precision. Even closer by, in-spirals of low-mass

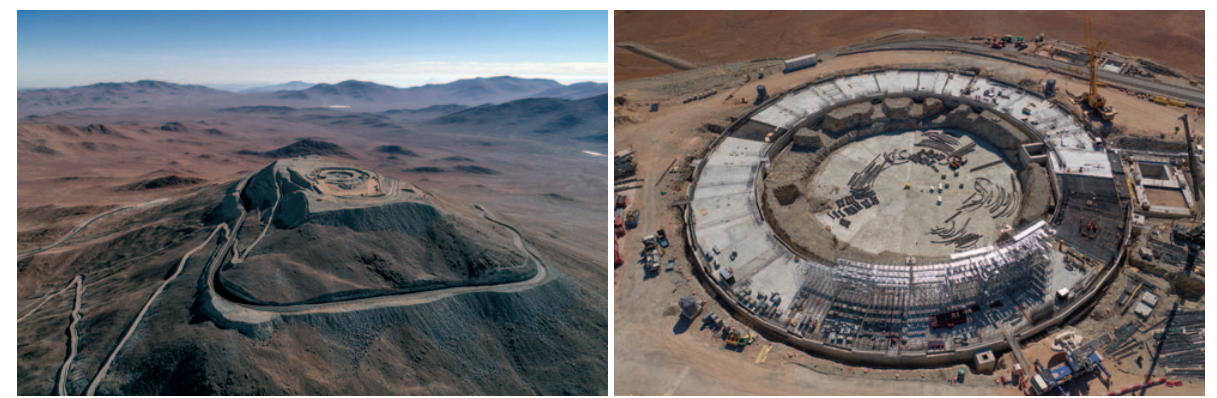

▲ Cerro Armazones in the Chilean Andes, with ESO's Extremely Large Telescope under construction (left) and a detailed view of the state of dome construction in January 2022 (right).

compact objects (white dwarfs, neutron stars and in particular black holes) spiralling into (super)-massive black holes will provide a census of massive black holes in galaxies. In the Milky Way and nearby galaxies, LISA will detect thousands of stellar-mass compact binaries: crucial information to test theories of binary evolution and compact object formation.

The development timeline for LISA is only very slightly later than Athena, ESA's flagship X-ray mission that resulted from our previous strategic plan, using national Roadmap funding. The synergy between the two will greatly enhance our community's opportunities for research: Athena has the study of black holes in the early Universe and the study of stellar-mass compact objects as

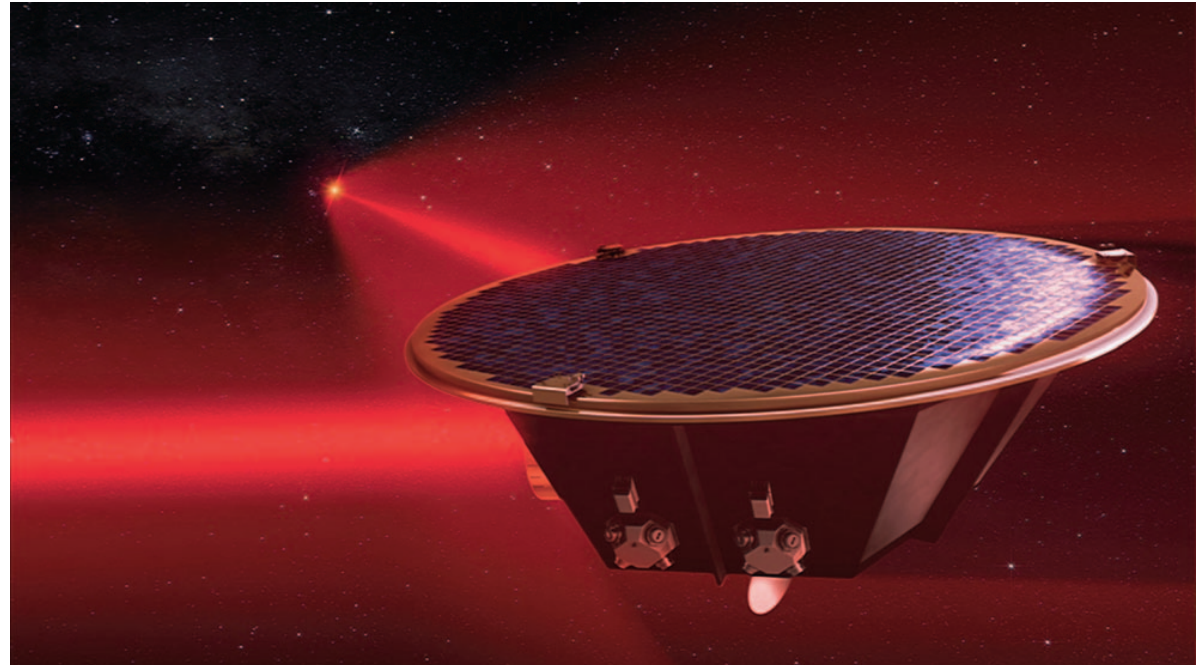

▲ *Artist's impression of one of the three satellites comprising the Laser Interferometer Space Antenna (LISA) and its laser connections to the other two. LISA is ESA's mission to explore low-frequency gravitational waves from sources throughout the Universe.*

We are very active in the international LISA consortium and have several leadership roles. SRON, together with Nikhef, TNO and industry, are developing key components such as the quadrant photodiodes for measuring the phase of the laser light and optical bench mechanisms and their electronics. These are at the heart of the mission and will give Dutch scientists an excellent understanding of the data. We will get preferred access to the data by contributing to the instrumentation, ensuring our participation in the first science with the mission and expert knowledge of the data for the whole community. Additional hardware investments are needed from the national Roadmap and for TNO via NSO or ESA. The physics community is leading the Einstein Telescope at the same time, resulting in close collaboration to make use of the technological overlaps and scientific synergies between these two important facilities.

key science goals and will offer a very complementary view of them to LISA.

**Delivering SKA and future radio instrumentation**
At the beginning of this decade, our radio astronomy is thriving:
- We are building an upgrade to LOFAR called LOFAR2.0, to greatly increase its sensitivity and data rate, due to go online in 2024. LOFAR has passed step 1 of becoming an ERIC and is due to complete its transformation into an ERIC by late 2022 or early 2023.
- In 2019 the Netherlands became a founding member of the Square Kilometre Array (SKA, an SP2011 top priority), the worldwide effort to build the largest radio telescope on Earth. After several decades in planning, phase 1 of the SKA is now being built in Western Australia and South(ern) Africa.

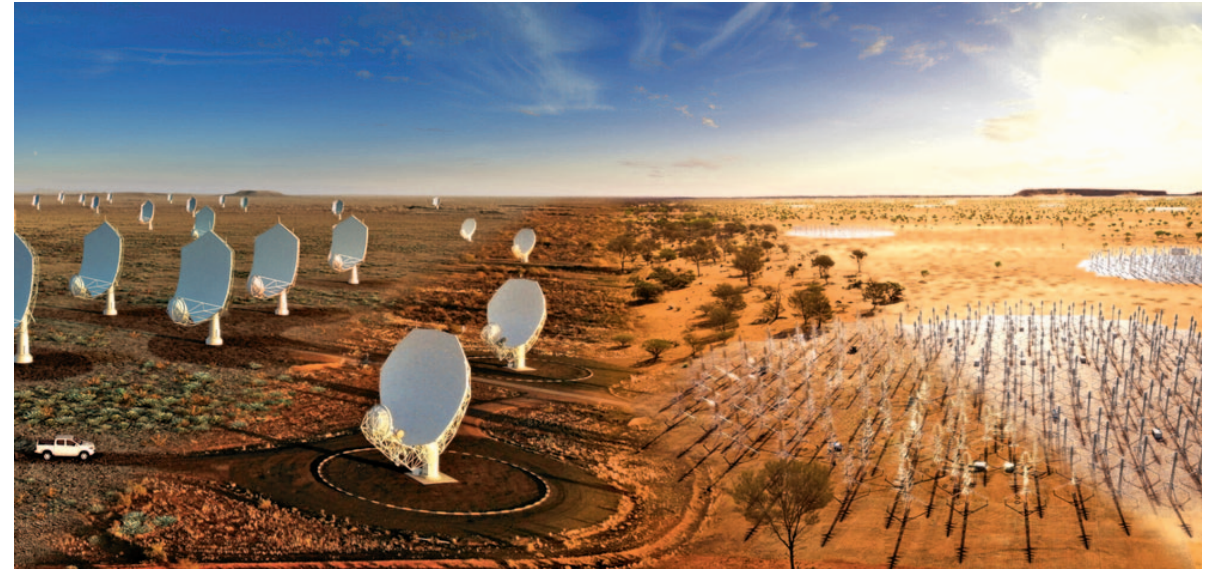

▲ *By the middle of the decade, construction of the Square Kilometre Array will be well underway in South Africa (left) and Australia (right).*

- JIV-ERIC, hosted in Dwingeloo, is a premier Very Long Baseline Interferometry expertise centre, which can combine the power of SKA with that of global networks of receivers into the highest-precision radio facilities.

Radio astronomy spans a factor million in wavelength, from sub-mm to the ionospheric cutoff, and we need multiple receiver and antenna technologies to cover that full range. We need to pursue several technology lines for them, from the tiny ultracold detectors of ALMA, developed from SRON technology in Groningen, to centimetre and metre-wave technology at ASTRON. There are also commonalities and synergies between them, such as the need to phase up receivers into multi-element interferometers, overcoming ionospheric and atmospheric disturbances.

To oversee and coordinate our strategy over this wide range, we are advised by the Netherlands Radio Astronomy Advisory Committee, NRAAC. It sees us very well positioned for a strong role over the full wavelength range, using ALMA, SKA and its precursors, and LOFAR. In the latter half of the coming decade, the centimetre- and metre-wave efforts will need significant investment to retain our leadership role. The most natural route now appears to be the SKA's development programme, by investing in a centimetre-wave (GHz frequency) facility in the Netherlands, both as elements in global VLBI with SKA and as a technology development platform for SKA upgrades. An alternative would be an upgrade of LOFAR2.0, capitalising on its complementary capabilities to SKA.

In addition to the instrument technology development, radio astronomy is particularly strongly computer driven. We deal with very large data volumes here and need very large amounts of processing power to transform our raw data into useful science-ready products. For this reason, we have invested heavily in developing software pipelines, and ASTRON are now developing (with national partners such as Nikhef) a Science Data Centre that can give Dutch radio astronomers a head start in exploiting the current and future radio arrays. ASTRON's Science Data Centre will also act as an SKA regional data centre to be the gateway of our community into SKA science.

### 3.2 FINANCIAL ROADMAP

Our priorities in Ch.3.1 translate into a financial roadmap for Dutch astronomy for the coming decade. The waterfall diagram shows the total investment in astronomy over the period 2017-2030 at ASTRON, SRON, the university NOVA institutes, facilities, and exploitation required for realising the ambitions in this Strategic plan.
Next, we show the funding status of the required

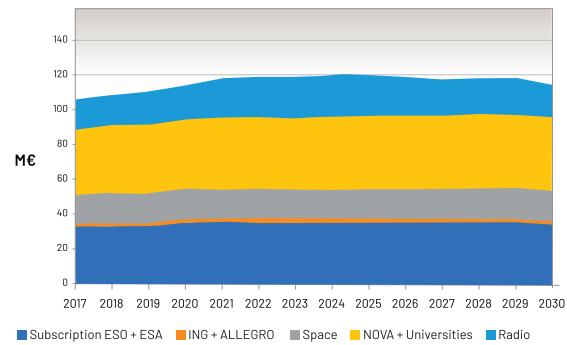

▲ *Total investment in astronomy in the Netherlands over the period 2017-2030. Numbers up to and including 2019 are actual expenditures. Numbers for the period 2021-2030 are forecasts in 2020 Euros, required to realise the plans presented here.*

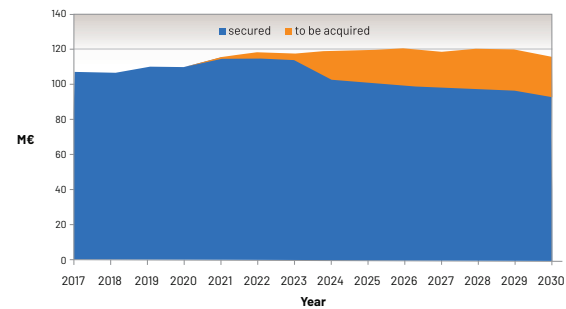

▲ *Funding status of the required investment in NL astronomy shown in the waterfall diagram. The 'to-be-secured' slice covers NWO Research Infrastructure and Large-Scale Research Infrastructure grants (including ELT instrumentation, radio astronomy, LISA) that are to be submitted/funded, the current funding horizon of 2023 for NOVA, and the continuation of the partnerships in Allegro, ING and JIV-ERIC.*

investments in NL astronomy. All budgets up to and including 2019 are actual budgets, not projections. The total budget over 2021-2030 is 1182 M€, of which the majority is secured, either through base budgets, extrapolation of research grants or through already allocated project financing. The fraction of the investment budget that remains to be secured is 13% of the total investment in astronomy. Note that the forecasted budgets are all in 2020 euros and not corrected for inflation.

We have been very successful in leveraging the base budgets of SRON and ASTRON and the university institutes over the period 2011-2021, more than doubling them with regional, national, and European grants and prizes for research and instrumentation. The same is true for the NOVA top research school funded since 1999. The budgets presented here assume a continuation of this success. These external resources are, however, additive: they can only materialize if the underlying foundation is sound and above a critical mass. In all, we can identify the key components of our funding in the following paragraphs.

### 3.2.1 BASE BUDGETS AND INTERNATIONAL CONTRIBUTIONS

- National contributions to the international treaty organisations ESO and ESA. The ESO contribution shows an increase until 2029: while 60% of the ELT construction costs (1B€) will come from the regular ESO budget and from the contribution of new member states, 40% comes from the existing members in the form of a special contribution, with the Dutch share 1/20th of that. The ESA science budget is assumed to be flat after a slight increase in 2020.
- Base budgets of the university institutes and SRON are assumed to be flat over the period 2021-2030. The ASTRON base budget also includes the costs for operating the LOFAR telescope and the Dutch contribution to the SKA.
- National contributions to the ING, Allegro, and JIV-ERIC are assumed to be flat for the coming decade, as needed given the expected level of activity of the WEAVE surveys and the ALMA and EVN science that these facilities support.

### 3.2.2 NOVA

Funding for the NOVA top research school was extended for five years in 2018, as part of the Bonus Incentives Scheme. Stable long term NOVA funding beyond 2023 is a top priority and is part of the 'to be secured' budget shown in the figure. A `sector plan' for astronomy has been proposed to OCW, as part of the sector plan 2022 for the science sector; at 7M€/y it will cover the instrumentation group and the data-science initiative (sect. 1.3). Funding for the research networks that have contributed so much to NOVA's impact (1.5M€/y) will be sought from elsewhere.

### 3.2.3 FLAGSHIP PROJECT COSTS

To realize our ambitions with respect to the ELT, SKA and LISA, we need dedicated long term project funding. We note that all these projects are on the National Roadmap. At the request of the Permanent Committee for Large Scale Research Infrastructure, astronomy and high energy physics have jointly established an integrated roadmap with priorities and phasing for the investments to be submitted to the National Roadmap calls, in which we included these three. ELT instrumentation requires funding for the manufacturing, assembling, integration and verification phase of METIS, Phase B studies until construction for MOSAIC and long-term R&D for EPICS. A proposal for funding has been submitted in the Roadmap Call 2021. The construction of SKA is largely funded until 2030 but requires additional funds for the operations phase and a data science centre. LISA requires funding in 2023 that has been requested as part of a joint astronomy-particle physics gravitational wave proposal to the Roadmap Call 2021.

- ELT-instrumentation: To realize the instrumentation programme for the ELT 16M€ in Roadmap funding will be needed. The finances for this NOVA-led programme are secured until 2023.
- LISA: 12 M€ is needed as of 2023 for a Dutch participation in this ESA mission as co-I.
- SKA: For the operations phase beyond 2030, 3-4 M€/yr is required. To set up and run a Dutch node of a global network of SKA Regional Centres, ~1-2 M€/yr will be needed from 2026.

### 3.2.4 NATIONAL AND INTERNATIONAL GRANTS

Regular funding lines for instrumentation provide the required funding at different stages and levels of instrumentation development.

- NWO Instrumentation: regular funding lines for instrumentation include the NWO competitions of NWO Research Infrastructure: national consortia (NWO-WI) and NWO ENW-M-Invest (Medium). The plans presented here represent an average of one Large astronomy project per call (every two years) with a budget of M€ 2.5 per project. There is a funding gap between ENW-M and NWO-Large. From astronomy it is desired to change the upper-limit of ENW-M investment grants to a level of at least 1.5 M€. With ever tighter university budgets, the financing of research (theoretical, numerical, and observational) depends more and more on grant competitions and prizes. The financial roadmap is based on continued success in acquiring funding from such channels. A rolling grant fund, that has been proposed to OCW, can reduce proposal pressure on the science system.
- NWO research grants include the NWO Open Competition and the NWO Talent programme that is a major factor in the continuing rejuvenation of Dutch astronomy by providing a strong backing for new faculty. To sustain these at the present level, NL Astronomy requires about 10 PhDs or postdoctoral positions per year through the ENW-OC-M, one ENW-OC-XL consortium grant every two years, as well as one Vici, 2 Vidi, and 4 Veni grants per year.
- The NWA has provided an impulse to some interdisciplinary work involving astronomy (sect.1.3), but due to the difficulty of engaging industry in astronomy research (as opposed to our success in engaging industry in our instrumentation effort, see sect.2.5), the interdisciplinary programmes of NWO have been a better match to our needs in this area.
- ERC Grants: The ERC Starting, Consolidator and Advanced Grants are an important funding line, where Dutch astronomy is doing exceptionally well (see Sect. 2.5). They fund both research and focused instrumentation projects; one of each of these per year should be sustainable based on past track record. Further ERC programmes, such as the Synergy Grant Scheme provide opportunities for larger (international) projects in particular in inter-disciplinary fields (a recent successful example is BlackHoleCam, yielding the images of black holes in M87 and our own Milky Way).
- Spinoza awards: The Spinoza awards (2.5 M€) are the most prestigious science awards in the Netherlands and not a grant competition. Seven Dutch astronomers have won the award since its introduction in 1995. Spinoza awards are used for a mix of research, instrumentation development and outreach.

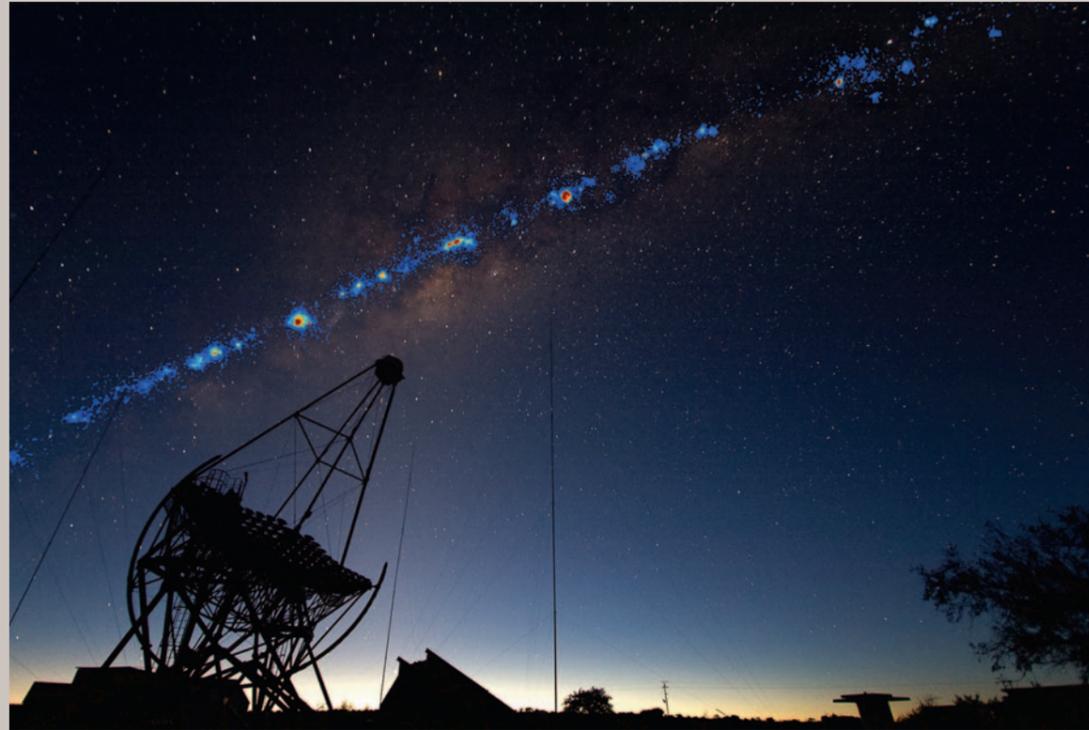

▼ *The central regions of the Milky Way over the HESS TeV gamma-ray telescope in Namibia. The colours indicate the sources of high-energy gamma rays in the plane of the Milky Way as imaged with HESS.*

F. ACERO & H. GAST

# 4 FROM THE PAST TO THE FUTURE

A strategic plan always involves assumptions about the future, and in the dynamical environment of modern astronomy, it is not a given that those are reliable, nor what its best timescale would be. Our choices of a decade for the planning period and of the appropriate ambition level are based on past experience. In this section, we provide a reality check on our strategic planning process and mention some instances of where we already need to look ahead further than the next decade to remain at the forefront of worldwide astronomy. Overall, the results of this exercise are very reassuring: we realised most of the priorities we set, and the science harvest from the facilities we invest in are quite rich, both in expected and unexpected ways.

## 4.1 RETROSPECTIVE OF 2011-2020

We briefly review the previous decadal period, both in terms of what we predicted in the decadal plan before (2001-2010), and in terms of how its priorities fared. First of all, such a look back effectively over 20 years emphasizes the long time scales in our field. An extreme example is the Dutch JWST/ MIRI participation, a priority of the 2001-2010 decadal plan. The instrument was launched in late 2021 and its first science harvest is being published at the same time as this decadal plan. Activities at strategic level for the period 2011-2020 consisted of the commissioning and science harvesting of major facilities that resulted from the 2001-2010 strategic plan, and seeking to secure continuity of NOVA, and membership of and funding for the next major initiatives.

### 4.1.1 SCIENCE HARVEST OF THE TOP FUNDING PRIORITIES OF 2001-2010

The facilities proposed in the 2001-2010 decadal plan were ALMA, VLT, LOFAR, Herschel/HIFI, and JWST/MIRI. They were indeed all funded and built, reasonably within the time scales and budget envelopes predicted for them (with the above-mentioned exception of JWST/MIRI, of which we hope to see the first harvest in late 2022). Each of these yielded results that one can fairly state even exceeded expectations and that cannot be done justice in the brevity of this report – we refer to the periodic activity reports of NOVA, SRON, and ASTRON for more extensive descriptions. We cannot, however, resist mentioning some highlights led by Dutch astronomers.

Supported by the national expertise centre Allegro, ALMA made detailed images of the discs around young stars in which planetary systems are forming. ALMA was also a key facility in creating the Event Horizon Telescope in the current decade, which was used to make the first picture of a black hole. This picture has been seen by a significant fraction of the world's population and was awarded the Breakthrough prize. Delicate measurements with the VLT enabled an initial charac-

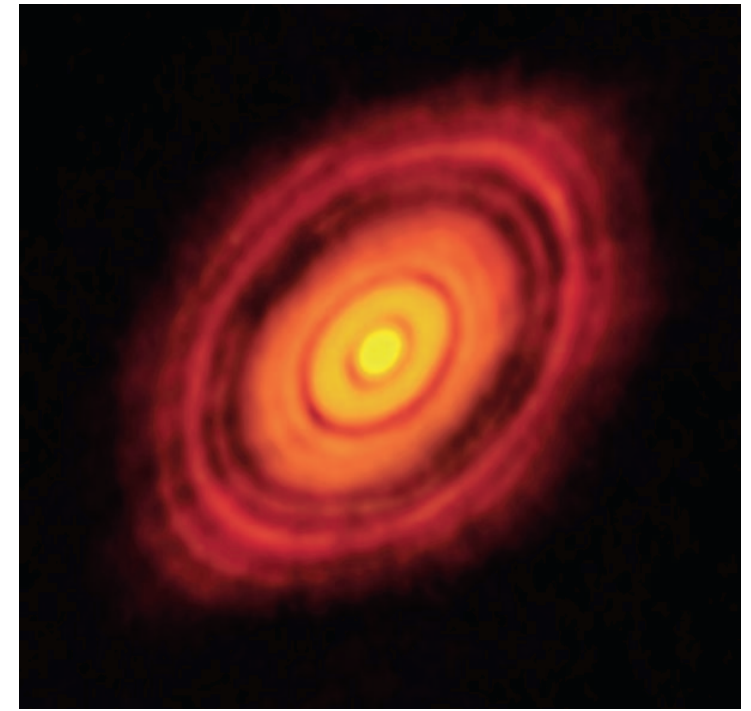

▲ *A planet-forming disc around the star HL Tau observed with ALMA. The dark rings mark absence of material, likely caused by the formation of planets in the disc.*

terisation of exoplanets and atmospheres. For the first time, the rotation rate of an exoplanet – Beta Pictoris b – was measured: its day lasts only eight hours. Making maps of the low frequency radio sky at unprecedented quality, LOFAR witnessed galaxy clusters colliding and forming the largest particle accelerators in the Universe.
A highlight of the Herschel/HIFI instrument was the study by of the role, origin and fate of water during star formation.
These top facilities in which we play a leading role benefit us in another important way: they are

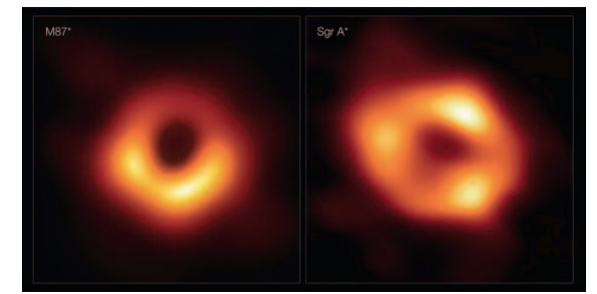

▲ *Images of two supermassive black holes made by the Event Horizon Telescope. On the right is the black hole in the centre of our Milky Way, and on the left is the 1000 times more massive black hole in the centre of the galaxy M87. It is about 1000 times bigger, but also 1000 times further away, and therefore appears the same size on the sky (EHT collaboration, 2019, 2022).*

open to others, and in exchange, we get to access facilities we did not play a significant construction role in (but identified as very desirable in the 2011-2020 plan). Sometimes this results from completely open-skies policies, but sometimes also from leveraging our 'own' facilities to collaborate with others, since many great discoveries are the result of combining results from multiple techniques. In this way, we played leading roles in many more of the top science results of the decade than we could from only the instruments we lead. Examples are the discovery of the first gravitational waves from massive merging black holes, and from the first binary neutron star merger, GW170817, and unveiling its connection to gamma-ray bursts and the formation of the most massive elements. Another example is the unravelling of the mystery of Fast Radio Bursts, by discovering that some of them repeat, and finding that they originate in somewhat distant galaxies. ESA's Gaia satellite measured the velocities and distances of about 1 billion stars; we used these these to discover that our Galaxy cannibalised another massive galaxy 10 billion years ago.

### 4.1.2 FATE OF THE TOP FUNDING PRIORITIES OF 2011-2020

The top funding priority in the 2011-2020 decadal plan was continued, and structural, funding for NOVA, for the reasons explained in Ch.2 and 3. We did succeed in extending its funding for further 5-year terms twice, which was very welcome. However, structural funding has still eluded us and it is as urgent as ever, so it is once again our top prior-ity. The major facilities proposed in the plan have also all been funded but are still being built. At the outset, these were leading an instrument for ESO's Extremely Large Telescope, participating in the global next-generation radio telescope SKA, and participating in the far-infrared space mission SPICA. The third of these had to be deprioritised in the midterm update, which has a severe effect on a long and successful line of far-IR instrument development in the Netherlands. It was replaced with the X-ray mission Athena. For all three eventual top facility priorities we have now secured funding.

**ELT:** The Extremely Large Telescope (ELT) is being built by the European Southern Observatory. With a 39-metre diameter mirror, it will be the largest optical-infrared telescope ever built. The ambition was to be PI of one of the instruments and this has been secured with NOVA taking the lead on the infrared instrument METIS. MICADO will provide a first-light capability for diffraction limited imaging at near-infrared wavelengths. NOVA is developing several subsystems for this. ELT is expected to begin science around 2027.

**SKA:** In 2019 the Netherlands became one of the founding members of the SKA Observatory, a new Intergovernmental Organisation that will lead the project. Construction of the SKA in South Africa and Australia started in June 2021 and the Netherlands will receive contracts that focus on delivery of SKA-low signal transport, the central signal processor and advanced algorithms for calibration and imaging. A Science Data Centre with associated software and hardware expertise to provide access to SKA data will also be set up. SKA is expected to begin science around 2027.

**Athena:** SRON obtained Roadmap funding for its co-PI role in the X-IFU instrument for Athena. This X-ray flagship mission by ESA, slated for launch in 2034, will enable SRON and the Dutch scientific community to continue its frontier role in X-ray studies of the extreme Universe.

Each of these key facilities builds on a history of over 50 years of expertise in science and instrumentation in its area and addresses a range of science goals with interesting overlaps, both with each other and with existing facilities, so their impact as a package is greater than the sum of their individual capabilities. We also see that two of three will indeed start science operations in the decade after the one in which they were selected, albeit a few years later than initially predicted; Athena provides another example of a longer timeline, with science operations starting in the middle of the next decade. Naturally, as new observatories and instruments come online, some venerable work horses also retire. We have scaled down our activity in some smaller optical telescopes, and after 50 years of frontier service, the Westerbork Synthesis Radio Telescope is ending active operations.

### 4.1.3 SEIZING OPPORTUNITIES AND SURPRISES

Even our expectation of surprises showing up can be verified in hindsight: in the past decade we seized on some opportunities that we did not predict in our strategic plan but proved to have high scientific impact. Fast radio bursts were only known as a single event in 2010, possibly even attributable to an instrumental error. Now they are a well-established phenomenon, likely some type of super-pulsars in distant galaxies, which our expertise in radio astronomy and in rapid counterpart identification enabled us to contribute to in a major way. The breakthroughs in gravitational-wave astronomy had been long in the making, but their rapid rise to prominence in the middle of the decade still came somewhat unexpectedly, and again we found ourselves in a strong position to study the electromagnetic counterparts with our experience in transient hunting. Our involvement in the NASA mission NICER, devoted to studying the dense matter equation of state, was requested because of our expertise, without investment in the instrument.

## 4.2 FORWARD LOOK TO 2030+ FACILITIES

It is evident from the previous sections that the timescales for addressing some of the most fundamental questions in astrophysics have become very long and go beyond the 10-year horizon of a single Strategic Plan. At the same time, we also notice how through the availability of new instrumentation, facilities, and modelling, sometimes unforeseen exciting new venues of research open. We therefore need to keep a careful balance between long-term plans and flexibility to seize new opportunities that may arise from new discoveries.

The simplest questions are often the most difficult to answer. Questions such as Are we alone? Why do galaxies and stars move so fast? How will

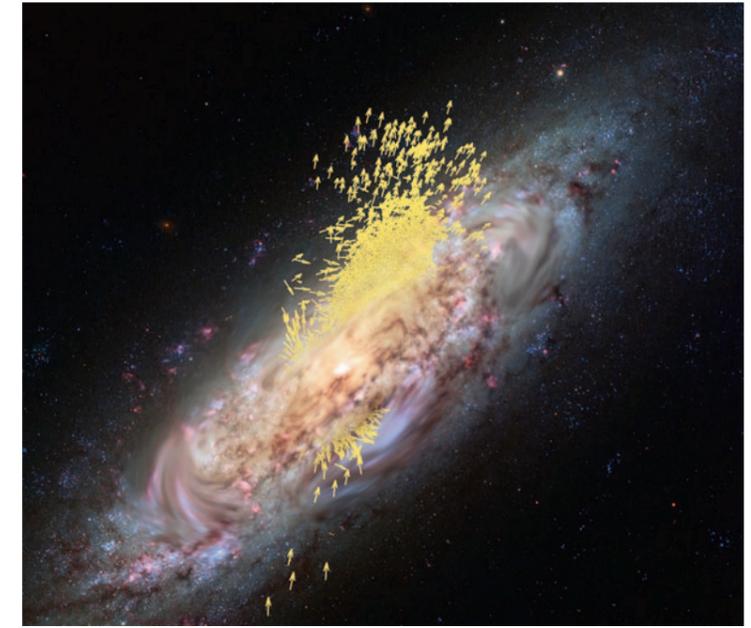

▲ *From the positions and velocities of stars measured by ESA's Gaia satellite, we infer that our Milky Way cannibalised another fairly massive galaxy about 10 billion years ago (Helmi et al. 2018).*

the Universe end? are fascinating and perplexing but turn out to be very challenging as well. We therefore need ever more complex facilities to address them, and by now their complexity is extreme. As a result, the timescales for their conception, design and construction have grown beyond the 10-year horizon.

We therefore already know some of the facilities, instruments and questions that will play a role in the 2030s. Such instruments and telescopes are in different phases of their development, but in all cases, they have been adopted by their respective agency. Amongst these projects, we find for the ELT instrumentation programme MOSAIC, and towards the end of the next decade, EPICS, the instrument aiming to directly image rocky exo-planets in the habitable zones of stars. Other projects that fall in this category are the ESA-led Athena and LISA. We therefore need to extend the cycle of conception, planning, building, and using them over multiple decadal plans, and invest significantly in R&D phases of facilities and missions that may be used in the 2040s and 2050s.

To anticipate and plan that long timeline, ESA recently carried out an exercise known as Voyage 2050, with the goal of identifying the important science questions and space missions needed to address those in the timeframe 2035-2050. At the end of 2019, the community was invited to submit ideas, and based on these a so-called Senior Committee has provided advice to the Science Programme Committee of ESA in the spring of 2021. The three very compelling science themes driving the largest missions are: I) moons of the giant Solar System planets, II) temperate exoplanets or the Galactic ecosystem, and III) new physical probes of the early Universe. For some of these themes, several options still need to be weighed in consultation with the community to eventually come to mission selection. Again, for the exceedingly complex missions that can address such exciting science we will need significant R&D development, and in practice the horizon will extend beyond 2050. This offers opportunities to participate in different ways and at different stages in such a mission, and we should be prepared for that by having the right technology developments, i.e., those that align the most with the science we wish to pursue.
On a different scale, there will also be smaller opportunities and the possibility of contributing to, e.g., US-led missions. An example is the UV-universe, which bridges many different topics in astrophysics from the atmospheres of planets to supernova explosions to the distribution of gas in and around galaxies and their evolution.

A firm desire of the astronomical community is to be able to "image" distant objects and to resolve their properties. An outstanding technique for this is via interferometry, and numerous ideas have been put forward to ESA for the use of this technique in different wavelengths (from the X-ray, FIR, submm to the radio) in space. This is relevant for understanding planets as well as for, e.g., the workings of supermassive black holes. ALMA has demonstrated the strength of interferometry in some of these fields, and we may expect that it will continue to do so in the next decade. Significant upgrades to the observatory are expected to occur in or around the 2030s.

Regarding ESO, it is clear that the cost and effort of building the ELT will prevent the organization from embarking on any new large programme at least until the first-light instruments have been commissioned. The expectation is that in the second half of the current decade, ESO will consult the community as to what to do next, with a horizon beyond 2035. In the context of the VLT in the 2030s, we also expect smaller projects to be developed. For example, with the success of the Gaia mission, spectroscopic follow-up work is expected to be intense in the coming decade, but may well extend to the 2030s, possibly with new dedicated facilities.

Similarly for the radio, although phase-1 of SKA may be completed around 2027, we expect further developments to follow between 2030 and 2035. Anticipated plans include targeted upgrades to some of the digital components (e.g., the correlator) to stay up to date with advances in computing technology. Further options are to extend the telescope's capacity by deploying new cooled wide-field (focal plane array) receivers on the SKA-mid dishes in South Africa or allowing the detection of ultra-high energy cosmic rays through addition of buffers and real-time event detection capability. We will need to make strategic choices to maximise the science return in the next decades and to preserve our leading role in technology development for radio astronomy applications.

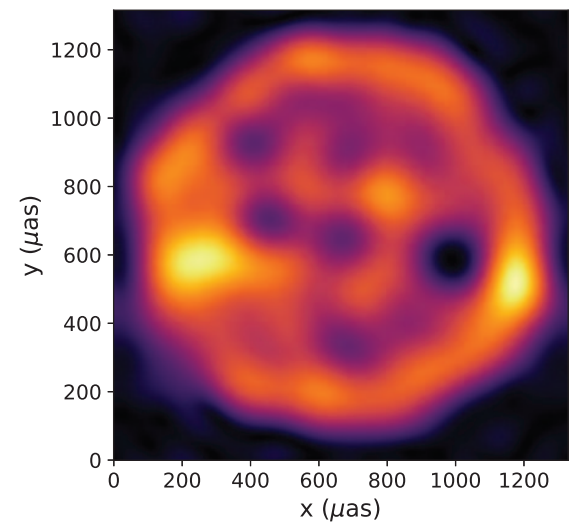

▲ *The image quality obtainable with a 1-m X-ray interferometer in space of the nearby (10 pc) Sun-like star AU Mic. Clearly visible are active regions and cooler region of the star, as well as the shadow of one of its Jupiter-like planets transiting in front of the star (right of centre).*

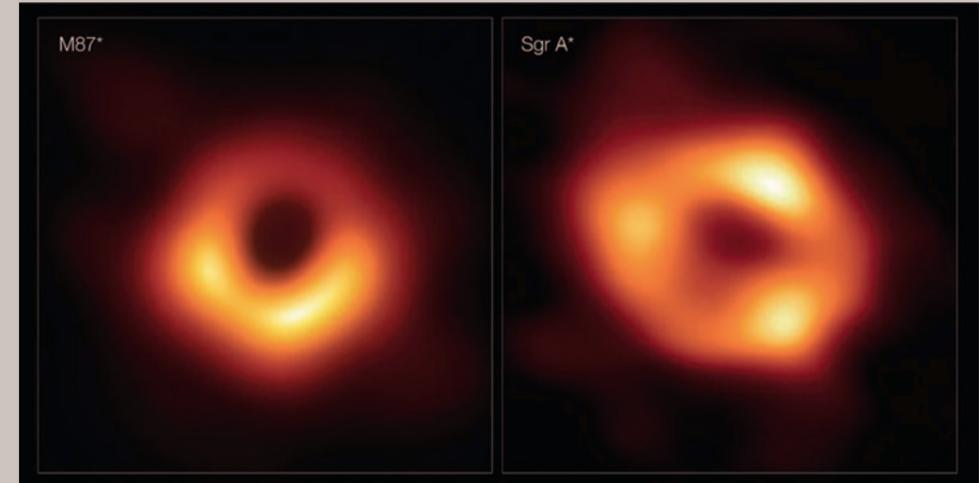

# CONCLUDING REMARKS

This strategic plan summarizes the ambitions of the Dutch astronomy research community for the next decade, as we continue our quest to study and understand our fascinating Universe.

We are excited to put in the effort and resources to strengthen our community, to design, build and work with new instruments and telescopes. We shall reach out to our partners at NWO and the government as well as scientists in related academic disciplines for the vital support that will be needed to make our plans a reality.

Ten years hence, we hope to be able to look back at yet another successful decade of Dutch astronomy. That success will be defined by our findings on habitable planets in other solar systems, the very earliest stars and galaxies, spacetime ripples and outbursts from colliding objects, and some unplanned surprises; and also very much by the inspiration those findings will provide to broad audiences.

# LIST OF ACRONYMS / GLOSSARY

**4MOST** – 4-metre Multi-Object Spectroscopic Telescope, future ESO optical spectroscopic survey instrument
**Allegro** – ALMA Local Expertise GROup, NL/European support centre for ALMA observations
**ALMA** – Atacama Large Millimetre Array, current ESO/global millimetre array
**Altair** – Diversity in STEM programme for elementary schools in Amsterdam
**AMT** – African Millimetre Telescope, future EHT extension in Namibia
**AMUSE** – Astrophysical MUlti-purpose Software Environment
**APERTIF** – APERture Tiles In Focus, wide field radio receivers for WSRT, being closed
**ASTRON** – Dutch Institute for Radio Astronomy
**Athena** – Advanced Telescope for High ENergy Astrophysics, future ESA X-ray mission
**BlackGEM** – optical telescope array for hunting gravitational-wave events, current at ESO
**CAN** – Committee for Astroparticle physics in the Netherlands (for monitoring and development of astroparticle physics)
**CCD** – Charge Coupled Device (an electronic optical detection device)
**CTA** – Cherenkov Telescope Array, future high-energy gamma-ray telescope
**DAN** – Dutch Astrochemistry Network
**EAGLE** – Evolution and Assembly of GaLaxies and their Environments (large-scale computer simulation project)
**EAS** – European Astronomical Society
**ECHO** – Expertise Centre for Diversity Policy
**EHT** – Event Horizon Telescope, current global millimetre-wave interferometer
**ELT** – Extremely Large Telescope, future 39-m ESO optica/infrared telescope
**EPICS** – ExoPlanets Imaging Camera and Spectrograph, 3rd-generation ELT instrument
**ERC** – European Research Council
**ERIC** – European Research Infrastructure Consortium
**ESA** – European space Agency
**ESO** – European southern Observatory
**ET** – Einstein Telescope, next-generation gravitational-wave observatory
**Euclid** – ESA mission under construction for exploration of dark matter and dark energy
**EVN** – European VLBI (Very Long Baseline Interferometry) Network
**eXTP** – enhanced X-ray Timing and Polarimetry mission, future China-led satellite
**Fermi** – current NASA gamma-ray satellite
**Gaia** – current ESA satellite for determining motions and distances of stars precisely
**GAMOW** – proposed NASA mission for exploring the earliest stars via gamma-ray bursts
**GR** – General Relativity (Einstein's theory of gravity)
**Herschel/HIFI** – former ESA far-infrared satellite and instrument
**HESS** – High-Energy Stereoscopic System, a gamma-ray telescope in Namibia
**HIRES** – HIgh REsolution Spectrograph, prospective 2nd-generation ELT instrument
**IAU** – International Astronomical Union, the global society of professional astronomers
**ING** – Isaac Newton Group, NL-UK observatory on La Palma
**INTEGRAL** – INTErnational Gamma-Ray Astrophysics Laboratory, current ESA gamma-ray satellite
**ISO-SWS** – Infrared Space Observatory - Short Wavelength Spectrograph, former ESA satellite
**IXPE** – Imaging X-ray Polarimetry Explorer, current NASA satellite
**JAXA** – Japanese Space Agency
**JIV-ERIC** – Joint Institute for VLBI in Europe ERIC, hosted at ASTRON
**JWST** – James Webb Space Telescope, current NASA infrared satellite
**JWST/MIRI** – Mid-InfraRed Instrument, partly NL instrument on JWST
**KiDS** – Kilo-Degree Survey, NL-led wide-field dark matter survey at ESO
**KM3NeT** – future neutrino telescope in the Mediterranean
**LIGO** – Laser Interferometer Gravitational-Wave Observatory, current US facility
**LISA** – Laser Interferometer Space Antenna, future ESA satellite
**LOFAR** – LOw Frequency ARray, current NL-led inernational radio telescope
**LOFAR2.0** – future upgrade version of LOFAR
**LVC** – LIGO-Virgo Collaboration, between the two eponymous observatories
**MeerKAT** – Meer Karoo Array Telescope, current South African radio telescope
**MeerLICHT** – single BlackGEM telescope in South Africa, co-observing with MeerKAT
**METIS** – Mid-Infrared ELT Imager and Spectrograph, NL-led 1st-generation ELT instrument
**MICADO** – Multi-AO Imaging Camera for Deep Observations, 1st-generation ELT instrument
**MOSAIC** – Multi-Object Spectrograph for Astrophysics, Intergalactic medium studies and Cosmology, 2nd-generation ELT instrument
**NAEIC** – Netherlands Astronomy Equity and Inclusion Committee
**NASA** – (US) National Aeronautics and Space Administration
**NICER** – Neutron star Interior Composition Explorer, current NASA X-ray satellite
**Nikhef** – National Institute for subatomic physics (of the Netherlands)
**NOVA** – Netherlands Research School for Astronomy, federation of the university astronomy institutes
**NSF** – (US) National Science Foundation
**NuSTAR** – Nuclear Spectroscopic Telescope ARray, current NASA hard X-ray satellite
**NUX** – future UV transient search telescope
**NWA** – (NL) National Science Agenda (Nationale Wetenschaps Agenda)
**NWO** – Netherlands Organisation for Scientific Research, funding agency
**NWO-EW** – (former) math/astro/comp.sci division of NWO
**NWO-ENW** – current natural science division of NWO
**OCW** – (NL) Ministry of Education, Culture, and Science
**PEPSci** – Planetary and ExoPlanetary Science programme, NWO-funded programme
**RFI** – Radio Frequency Interference, terrestrial signals perturbing radio telescopes
**RvdA** – Astronomy Council, Strategic coordination body of NL astronomy
**SEP** – Standard Evaluation Protocol for assessing NL university research
**SKA** – Square Kilometre Array, future global radio telescope
**SKA1** – phase 1 of SKA
**SPEX** – SRON-developed X-ray spectral analysis software (not to be confused with the English punk rock band "X-Ray Spex")
**SPHERE** – Spectro-Polarimetric High-contrast Exoplanet REsearch, ESO VLT instrument
**SPICA** – SPace IR telescope for Cosmology, once-planned ESA/JAXA satellite
**SRON** – Netherlands Institute for Space Research
**Swift** – current NASA gamma-ray burst and transient satellite
**THESEUS** – Transient High-Energy Sky and Early Universe Surveyor, proposed ESA gamma-ray burst satellite
**TNO** – Netherlands Organisation for Applied Scientific Research
**Virgo** – European gravitational-wave observatory
**VLBI** – Very Long Baseline Interferometry, combining distant radio telescopes for high resolution imaging
**VLT** – Very Large Telescope, current ESO facility at Paranal, Chile
**Voyage 2050** – ESA's forward planning programme for future missions
**VST** – VLT Survey Telescope, current ESO facility at Paranal
**WARP** – WAves, Rays and Particles, NWO-funded programme in astroparticle physics
**WEAVE** – WHT Enhanced Area Velocity Explorer, multi-fibre spectrograph for the WHT, one of the ING telescopes
**WHT** – William Herschel Telescope, current 4.2-m telescope in the ING
**WSRT** – Westerbork Synthesis Radio Telescope, current facility of ASTRON
**X-IFU** – X-ray Integral Field Unit, instrument on ESA's Athena satellite
**XMM-Newton** – current ESA X-ray satellite
**XRISM (XARM)** – X-Ray Imaging and Spectroscopy Mission, future JAXA/NASA X-ray satellite, previously known as XARM

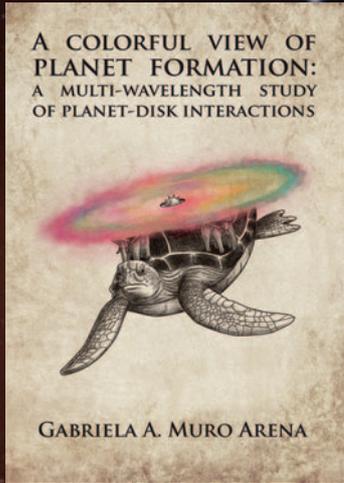 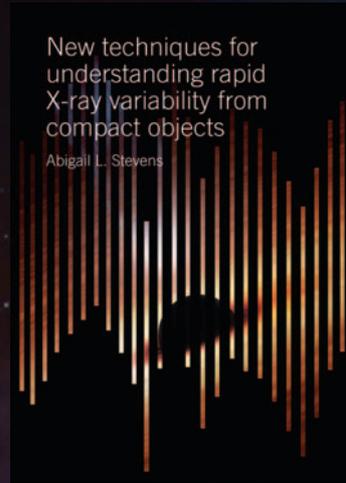 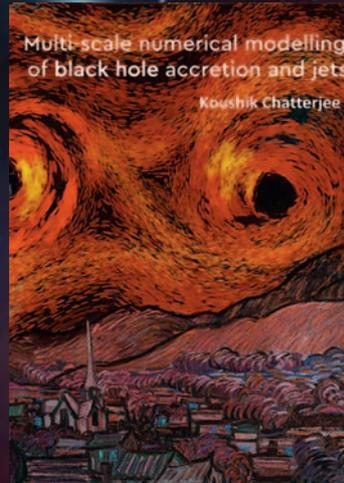 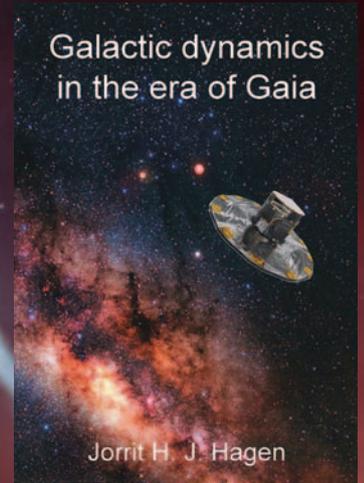
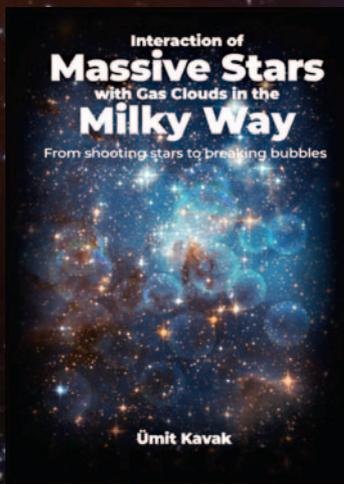 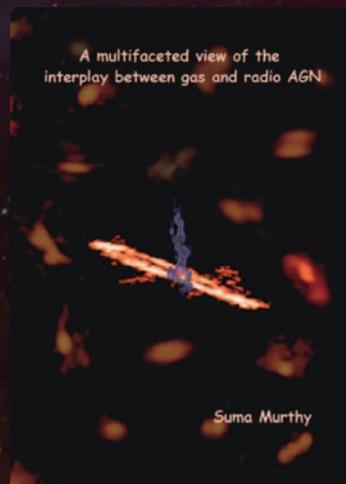 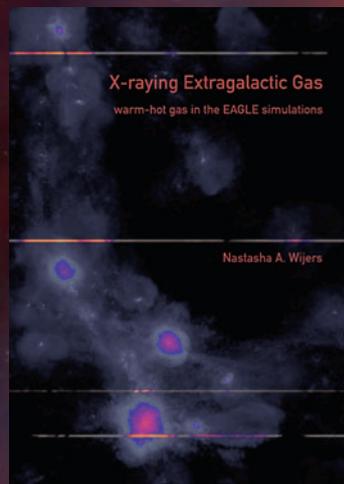 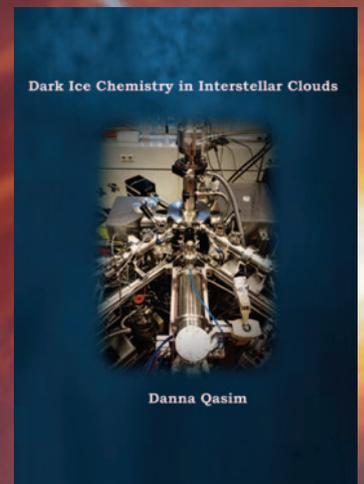
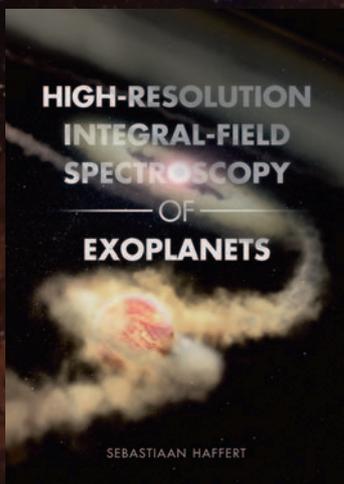 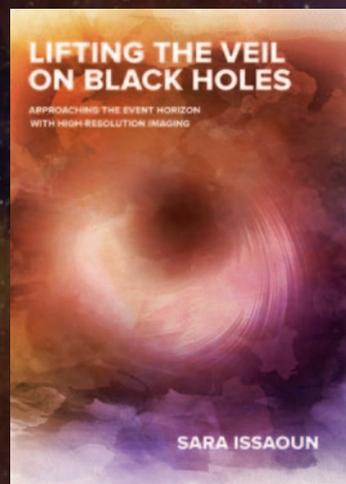 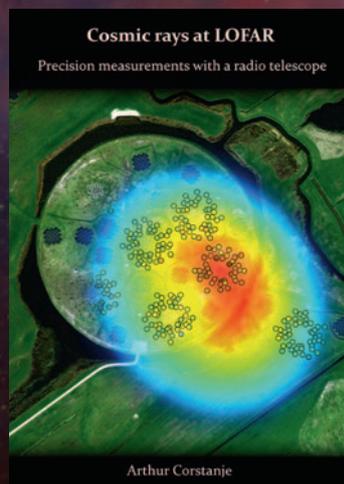 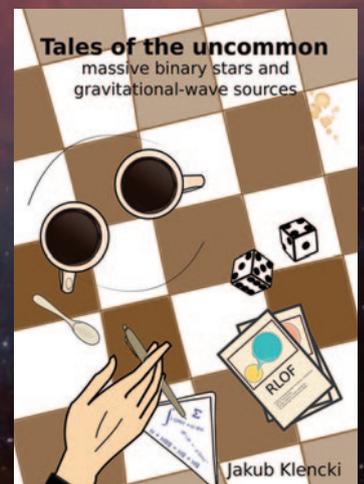

## PHD THESES IN ASTRONOMY

PHD theses in the past few years of the Netherlands,

showing the rich diversity of our scientific harvest.